\documentclass[fleqn,usenatbib]{mnras}

\usepackage{newtxtext,newtxmath}

\usepackage[T1]{fontenc}

\DeclareRobustCommand{\VAN}[3]{#2}
\let\VANthebibliography\thebibliography
\def\thebibliography{\DeclareRobustCommand{\VAN}[3]{##3}\VANthebibliography}

\usepackage{graphicx}	
\usepackage{amsmath}	
 
\usepackage{amssymb}	
\usepackage{xcolor}
\usepackage{ulem}

\newcommand{\simba}{{\sc Simba}}
\newcommand{\Msun}{M$_{\sun}$}

\newcommand{\gizmo}{{\sc Gizmo}}
\newcommand{\mufasa}{{\sc Mufasa}}

\newcommand\orcidicon[1]{\href{https://orcid.org/#1}{\includegraphics[scale=0.65]{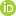}}}

\newcommand{\revision}[1]{#1}
\newcommand{\update}[1]{#1}

\title[Rapidly quenched galaxies]{Rapidly quenched galaxies in the \simba\ cosmological simulation and observations}

\author[Y. Zheng et al.]{
Yirui Zheng\orcidicon{0000-0001-7707-5930},$^{1}$\thanks{E-mail:yz69@st-andrews.ac.uk} 
Romeel Dave\orcidicon{0000-0003-2842-9434}, $^{2,3,4}$
Vivienne Wild\orcidicon{0000-0002-8956-7024}$^{1}$
and Francisco Rodr\'iguez Montero\orcidicon{0000-0001-6535-1766}$^{5}$
\\
$^{1}$School of Physics and Astronomy, University of St Andrews, North Haugh, St Andrews, Fife, KY16 9SS, Scotland, UK\\
$^{2}$Institute for Astronomy, Royal Observatory, University of Edinburgh, Edinburgh EH9 3HJ, UK\\
$^{3}$University of the Western Cape, Bellville, Cape Town 7535, South Africa\\
$^{4}$South African Astronomical Observatories, Observatory, Cape Town 7925, South Africa\\
$^{5}$Astrophysics, University of Oxford, Keble Road, Oxford OX1 3RH, UK
}

\date{Accepted XXX. Received YYY; in original form ZZZ}

\pubyear{2021}

\begin{document}
\label{firstpage}
\pagerange{\pageref{firstpage}--\pageref{lastpage}}
\maketitle

\defcitealias{rodriguez-montero2019}{RM19}

\begin{abstract}
\revision{Galaxies with little star formation are found to have quenched over a variety of timescales, which provides insights into the physical mechanisms responsible. Here we examine}
the population of rapidly quenched galaxies (RQGs) in
the \simba\ cosmological hydrodynamic simulation at $0.5<z<2$, and compare them directly to observed post-starburst galaxies in the UKIDSS Ultra Deep Survey (UDS) via their colour distributions and mass functions. We find that the fraction of quiescent galaxies that are rapidly quenched in \simba\ \revision{at $z=1$} is 
\revision{$59\pm3$ per cent,
contributing $48\pm5$ per cent to the total mass of the red sequence,
which is at the upper end of the $\sim$25--50 per cent derived from the UDS.}
A similar ``downsizing'' of RQGs is observed in both \simba\ and the UDS, with RQGs at higher redshift having a higher average mass. However, \simba\ produces too many RQGs at $1<z_q<1.5$ and too few low mass RQGs at $0.5<z_q<1$. Comparing colour distributions further suggest discrepancies in star formation and/or chemical enrichment histories, including an absence of short, intense starbursts in \simba. Our results will help inform the next generation of galaxy evolution models, particularly with respect to the quenching mechanisms employed.

\end{abstract}




\section{Introduction}
Galaxies in the local Universe display a strong bimodality in the colour–magnitude diagram \citep{Baldry_2004, Jin_2014}.
Most massive galaxies in the local Universe can be classified into two distinct groups: the ‘red sequence’ (quiescent, elliptical galaxies) and  the ‘blue cloud’ (star-forming, spiral galaxies). 
Observations have revealed that the total number of stars that live in quiescent galaxies (i.e. the total stellar mass), as well as the total number of quiescent galaxies, continue to increase since a redshift of $z\sim1$; in contrast, the stellar mass density of star-forming galaxies remains almost constant \citep{bell2004nearly,Ilbert2013,Muzzin2013}.
Quiescent galaxies have little in-situ star formation activity, thus, the steady growth of the red sequence implies a steady conversion of star-forming galaxies into quiescent ones. However, how and why galaxies shut off their star formation and build up the red sequence remains unclear.  

A natural idea is that the star-forming galaxies deplete their gas supply and the star formation is gradually quenched. The Universe's global star formation rate density has declined by a factor of $\sim10$ since $z\sim2$ \citep{Madau2014}, and as the number density of star-forming galaxies has remained relatively constant during the same time, this strongly indicates that typical star-forming galaxies are gradually decreasing their star formation rate over the past several billion years (Gyr).
However, morphological observations suggest that gas depletion may not be the dominant quenching mechanism. A blue spiral galaxy will not experience a morphological transformation when it is quenched by simple gas depletion, while observations reveal a strong correlation between quiescent galaxies
and a predominantly spheroid morphology at all redshifts 
\citep[e.g.][]{Bell2012, Lang2014,Bruce2014,Brennan2015, Ownsworth2016}.
Meanwhile, simple gas depletion can take a timescale of several Gyr, which makes it hard to explain the existence of quiescent galaxies in the relatively young Universe \citep[out to $z \sim 4$ or more, e.g.][]{straatman2014zfourge}.
Further analysis of the star formation histories (SFHs) of quiescent galaxies leads to growing evidence of the existence of rapid quenching mechanisms, alongside other relatively slow quenching mechanisms such as depletion of gas supply  \citep[e.g.][]{moutard2016vipers, Pacifici2016, Maltby2018,Rowlands2018GAMA,wu2018fast, Belli2019, Wild2020star}.

Many mechanisms may be responsible for the quenching of galaxies. They broadly fall into two categories, internal and external. Popular internal mechanisms to shut off star-formation are active galactic nucleus (AGN) feedback and morphological quenching. Models suggest that AGN can heat up the cold gas in the galaxies by thermal feedback and/or expel the cold gas by kinetic feedback, leading to a reduced gas content for star formation \citep[][]{dimatteo2005energy, choi2014consequences, Dave2019, zheng2020comparison}, 
although observations in support of this scenario remain indirect and/or subjective  \citep[e.g.][]{Best2005, Kaviraj2007,Smethurst2017, french2018clocking}.
The morphology of galaxies may likewise prevent star formation or accelerate gas depletion. Once a galaxy acquires a spheroid-dominated morphology, models suggest disc stabilisation may stop gas cloud fragmentation and therefore gradually halt the star formation \citep{Martig2009}.

Various external mechanisms may also be able to shut-off star formation and make galaxies quiescent, depending on the environment of the galaxies. Ram-pressure stripping can remove the cold gas content of satellite galaxies when they fall into a massive cluster \citep{gunn1972infall}; `strangulation’ prevents the galaxies from retaining a gaseous halo that is required to continually fuel the disc in the cluster environment \citep{larson1980evolution, balogh2000h}; fast encounters with other galaxies can disturb the star formation within the galaxies
\citep[`harassment’,][]{gallagher1972note, moore1998morphological}.
In less dense environments, mergers are more common, and might be responsible for quenching: the powerful starbursts triggered by a merger quickly consume the gas content and models suggest that the subsequent stellar feedback (and possibly AGN feedback) could violently heat and expel the remaining gas 
\citep[e.g.][]{barnes1992transformations, naab2003statistical,dimatteo2005energy,bournaud2005galaxy}. However, there is no obvious link between mergers and quenching in cosmological hydrodynamic simulations \citep[][]{rodriguez-montero2019,Quai2021} and, while it is challenging to link the two events observationally, circumstantial evidence indicates a lack of connection \citep[][]{Weigel2017,Ellison2018}. 

The lack of clarity from observations despite increasingly impressive datasets suggests a complex array of internal mechanisms are at play, even before environmental processes are introduced. Bulge-to-total mass ratio, absolute bulge mass, molecular gas fraction and the presence of an AGN are strongly correlated with reduced star formation activity in galaxies \citep[see e.g.][for some recent results]{bluck2014bulge, Zhang2021}. Both ``slow'' and ``rapid'' quenching processes likely play a role, with the balance of mechanisms depending on redshift, stellar mass and environment. The ``slow'' process is likely multi-faceted, with steady bulge build-up leading to disk stabilisation and reduced efficiency for star formation in surrounding disks, and concurrent build-up of black hole mass (from the same gas source that built the bulge) leading to energy input from the AGN preventing any further gas cooling in the most massive bulges \citep{bluck2014bulge}. Subsequent temporally disconnected mergers might then lead to elliptical-like morphologies \citep{Zhang2021}. On the other hand ``rapid'' processes appear to have more concurrent morphological transformation \citep[][]{Yano2016,Almaini2017,Maltby2018} and could be related to the extreme starburst events identified in the sub-millimetre \citep[][]{Wild2020star, Wilkinson2021}.

Fundamentally, the large number of galaxy properties which correlate with star formation activity in galaxies makes it difficult to disentangle causation from correlation without reliance on extensive simulations, which themselves are limited by resolution, size and input physics. Higher order observations, such as the timescales of the quenching events, and the previous evolutionary history from the stellar fossil record, are starting to play an important role in revealing the balance of quenching mechanisms responsible for forming today's quiescent galaxy population. 

Two mechanisms for rapidly quenching galaxies which are thought to operate on time scales shorter than a couple of hundred Myr are kinetic AGN feedback \citep[e.g.][]{kaviraj2011simple}, and ram-pressure stripping which operates on galaxy cluster crossing time-scales \citep[e.g.][]{abadi1999ram}. The resulting rapid truncation of star formation can be identified in the stellar populations of the galaxies.  O- and B-type stars explode as supernovae a few million years after the quenching event, leading to A- and F-type stars dominating the light of the galaxy. This results in little/no emission lines but strong Balmer absorption lines and Balmer break. The Balmer absorption features are stronger when the galaxies experience starbursts prior to the rapid quenching event.  These galaxies are usually called `post-starburst' galaxies (PSB).  

Traditionally, combinations of line indices are used to select rapidly quenched galaxies or PSB candidates, e.g. the H$\alpha$ emission line
\citep[e.g.][]{goto2003hdelta, quintero2004selection, balogh2005near}
or the [OII] emission line 
\citep[e.g.][]{dressler1983spectroscopy, zabludoff1996environment, poggianti1999star}
to quantify the weak or absent nebular emission, combined with H$\delta$ or $H\gamma$ absorption to quantify the Balmer absorption strength. Alternatively, the 4000\AA\ break may be used to estimate ongoing star formation rate, rather than nebular emission lines \citep{kauffmann2003stellar}. Since quiescent galaxies that host narrow line AGN can be excluded by an emission line cut, \cite{wild2007bursty} promotes a principal component analysis (PCA) based only on the stellar continuum around 4000\AA\ to select PSBs. When spectra are not available, broad band photometry can also be used, e.g. the rest-frame UVJ diagram used in \cite{Whitaker2012} and the principal component analysis (PCA) super-colour method introduced in \cite{Wild2014b}.

The contribution of the rapid quenching route to the growth of the red sequence has been a topic of research for the last decade.  \cite{wild2009post} used spectra from the VVDS survey to estimate that $\sim$40\% of the mass growth of the quiescent population at $z\sim0.7$ is contributed by PSBs, and \cite{Wild2020star} combine photometry and spectra from the UDS survey to estimate a value of 25-50\%\ at $z\sim1$ \footnote{\cite{Wild2020star} identify PSBs by their super-colours (read more about super-colours in \autoref{subsec:sca} for more about super-colours).}. \cite{Belli2019} estimate that rapid quenching accounts for $\sim20$\% of the growth of the red sequence at $z\sim 1.4$ and $\sim50$\% by $z\sim 2.2$\footnote{ \cite{Belli2019} identify fast quenched galaxies via the UVJ diagram.}.
\revision{\cite{tacchella2022fast} find that about 25\% of the quiescent galaxies in Keck/DEIMOS survey HALO7D are quenched within a short timescale of 500 Myr at $z_{obs}\approx0.8$. }
The importance of the rapid quenching route appears to decrease with decreasing redshift and may rapidly diminish at $z < 1$ \citep{wild2016,Rowlands2018GAMA}, and appears to also depend on stellar mass and environment. \cite{socolovsky2018} find a much higher percentage of $\sim70$\% of quiescent cluster galaxies with $9.0 < \log({\rm M/M_\odot}) < 10.5$ and $0.5<z<1.0$ have rapidly quenched, although this may not be true of all clusters \citep[][]{DeLucia2009}. There is a strong link with galaxy morphology, with rapidly quenched galaxies being highly compact \citep{Yano2016, Almaini2017, Maltby2018}
and compact galaxies at intermediate redshifts found to have faster quenching times than normal-sized galaxies \citep{wu2018fast, nogueira2019compact}.

Recently, the \simba\ cosmological simulation was found to produce a bimodality in quenching timescales \citep[][RM19 hereafter]{rodriguez-montero2019}, which provides a great platform to further investigate the contribution of rapid quenching to the growth of the red sequence. In this paper, we will combine the known star formation histories (SFH) of the model galaxies with mock photometric data to further investigate the rapidly quenched galaxies and figure out the relative importance of the different quenching routes. 
In Section 2 we describe the \simba\ cosmological simulation, how we identify galaxies 
and extract their star formation histories, the mock photometry and super-colour analysis.
In Section 3, we present our results and findings. Finally in Sections 4 and 5 we discuss and summarise the results. 

\section{SIMULATION AND ANALYSIS}
\label{sec:sim}

In this paper we study the \simba\ simulation where we have access both to the true star formation histories of the galaxies, and can build mock photometric datasets with which to compare to galaxies in the real Universe. We focus on the sample of 2623 galaxies that are quiescent at a redshift of $z\sim 1$ (snapshot 105 in \simba), in order to understand how they became quenched. In this section we describe the simulation and methods used.

\subsection{The \simba\ simulation}
\label{subsec:simba}
In this paper, we study the \simba\ cosmological hydrodynamic simulation \citep{Dave2019}, the next generation of the \mufasa\ simulation \citep{Dave2016}. Like \mufasa, \simba\  employs a forked version of \gizmo, a mesh-less finite mass hydrodynamics code \citep{hopkins2015new} based on Gadget-3 \citep{springel2005cosmological}. \simba\ uses a set of state-of-the-art sub-resolution prescriptions, including H$_2$ based star formation rate \citep{krumholz2011comparison}, supernova feedback, star formation-driven galactic winds, a chemical enrichment model that tracks 11 elements, radiative cooling and photoionisation heating. These models are slightly updated to match the recent theoretical and observation results (see \citealp{Dave2019} for more details).

The major updates of \simba\ compared to \mufasa\ are in two aspects.  Firstly, \simba\ includes on-the-fly dust production, growth and destruction. Secondly, \simba\ tracks the growth of super massive black hole (SMBH) accretion and associated feedback. \cite{Dave2019} describes the SMBH accretion model: Hot gas ($T >10^5$K) is accreted onto the central black hole via the standard Bondi–Hoyle–Lyttleton accretion mode \citep{bondi1952spherically}, while cold gas is accreted through the action of gravitational torques following the analytic sub-grid model of \citet{hopkins2011analytic}.
For feedback, \simba\ adopts two different SMBH feedback models, depending on the Eddington ratio. If the SMBH is in a high Eddington ratio state ($f_{\rm Edd}$>0.2), the SMBH will eject radiative AGN winds at velocities $v\sim 10^3$~km/s without changing the gas temperature; this is broadly similar to the AGN feedback model introduced in \cite{angles2017gravitational}.  As $f_{\rm Edd}$ is lowered, feedback transitions to a jet mode with the outflow velocity is strongly increased. The feedback becomes entirely jet mode when the Eddington ratio is lower than 0.02 and SMBH mass is higher than $10^{7.5}$ \Msun.
\cite{Dave2019} showed that massive quiescent galaxies are primarily quenched by the jet feedback from the SMBH at the galaxy centre.  Both feedback modes are ejected bipolarly, along the central region's angular momentum axis.  When full jets are one, a third AGN feedback mode via X-ray heating of central galactic gas turns on \citep[following][]{choi2012}; this mode is important for fully quenching galaxies.

We use the fiducial run of the \simba\ simulation suite, which has a box length of 100 comoving $h^{-1}$ Mpc and a minimum gravitational softening length of 0.5 comoving $h^{-1}$ kpc.
The simulation has $2\times1024^3$ particles, with equal numbers of baryonic and dark matter particles. The particle mass resolution is $9.6\times10^7$ \Msun\ for dark matter and $1.2\times10^7$ for the initial gas particles. In the fiducial run, the simulation begins at a redshift of $z=249$ and ends at $z=0$, thus simulating our present day Universe.
The first snapshot is output at $z=20$ and snapshots are then output with a time spacing that is comparable to galaxy dynamical timescales, which yields 151 snapshots in total. 
We are interested in the redshift range of $z=2$ to $z=0.5$, where the snapshots have a time spacing of $\sim$ 100 Myr and $\sim$ 150 Myr respectively. 
\simba\ takes a Planck Collaboration XIII (\citeyear{Planck2016}) concordant cosmology of $\Omega_m=0.3$,  $\Omega_\Lambda=0.7$, $\Omega_b=0.048, H_0=68$ km\,s$^{-1}$\,Mpc$^{-1}$, $\sigma_8=0.82$ and $n_s=0.97$.

We use the CAESAR\footnote{\url{ https://caesar.readthedocs.io/en/latest/}} package to analyse the galaxy properties by firstly grouping  stars
and interstellar medium (ISM) gas into baryonic galaxies via a 6D friends-of-friends (FOF) algorithm. The FOF adopts a spatial linking length of 0.0056 times the mean inter-particle spacing and a velocity linking length of the local velocity dispersion. A resolved galaxy has a minimum number of 32 particles, which yields a minimum stellar mass of $5.8\times10^8$ \Msun.

The galaxy properties in \simba\ are in good agreement with observations \citep{Dave2019}. A wide range of galaxy statistics are well reproduced, including the galaxy stellar mass function evolution, star formation rates, metallicities, dust properties, and black hole properties.
The good agreement between \simba\ and observations makes it a useful platform for investigating the specific physical processes driving galaxy evolution in more detail.

\subsection{Star formation histories and quenching timescales}
\label{subsec:SFH_qts}

Given the star formation histories (SFH) of galaxies, it is straightforward to measure the timescale over which they quench, and with simulation data this is possible. We identify galaxies to be quiescent if they have an instantaneous specific star formation rate sSFR $< 0.2/t_H(z)$, where $t_H(z)$ is the age of the Universe at redshift $z$ and sSFR is the ratio of instantaneous star formation rate (SFR) to the total stellar mass of the galaxy, sSFR$=$SFR/M$_*$,
\revision{
following \citetalias{rodriguez-montero2019} and \cite{Pacifici2016}.
Star-forming galaxies form a tight main sequence on the $SFR-M_*$ plane, whose formation timescales are typically seen to be smaller than the Hubble time at that redshift \citep[e.g.][]{Whitaker2012SFMS,fumagalli2014dead, speagle2014highly}.  This motivated \cite{Pacifici2016} to parameterise the evolution of the quiescent threshold as sSFR=$0.2/t_H(z)$, meaning that quenched galaxies are defined to have a formation timescale of $>5\ t_H$, and therefore lie significantly below the main sequence at all relevant redshifts.
}

To calculate the SFHs of each galaxy, we use the instantaneous SFR of the galaxy at each snapshot, tracking back through the most massive progenitor. In the case that there are multiple galaxies in the preceding snapshot due to merger events, we choose the most massive one as the progenitor and follow this progenitor back in time. By such a process, we track backwards the most massive branch of the merger tree of the quiescent galaxies. In some rare cases, we choose to select the longer branch rather than the most massive progenitor branch. This is due to limitations of the halo finder program, when the most massive progenitor has no progenitors, while the second most massive does have. In these cases, we chose the second most massive progenitor to allow us to track back the target galaxies over a longer time range. Fortunately, such cases happen in only about 2\% of the sample, so we do not anticipate any impact on our statistical results due to using the longer branches.

\begin{figure}
	\includegraphics[width=\columnwidth]{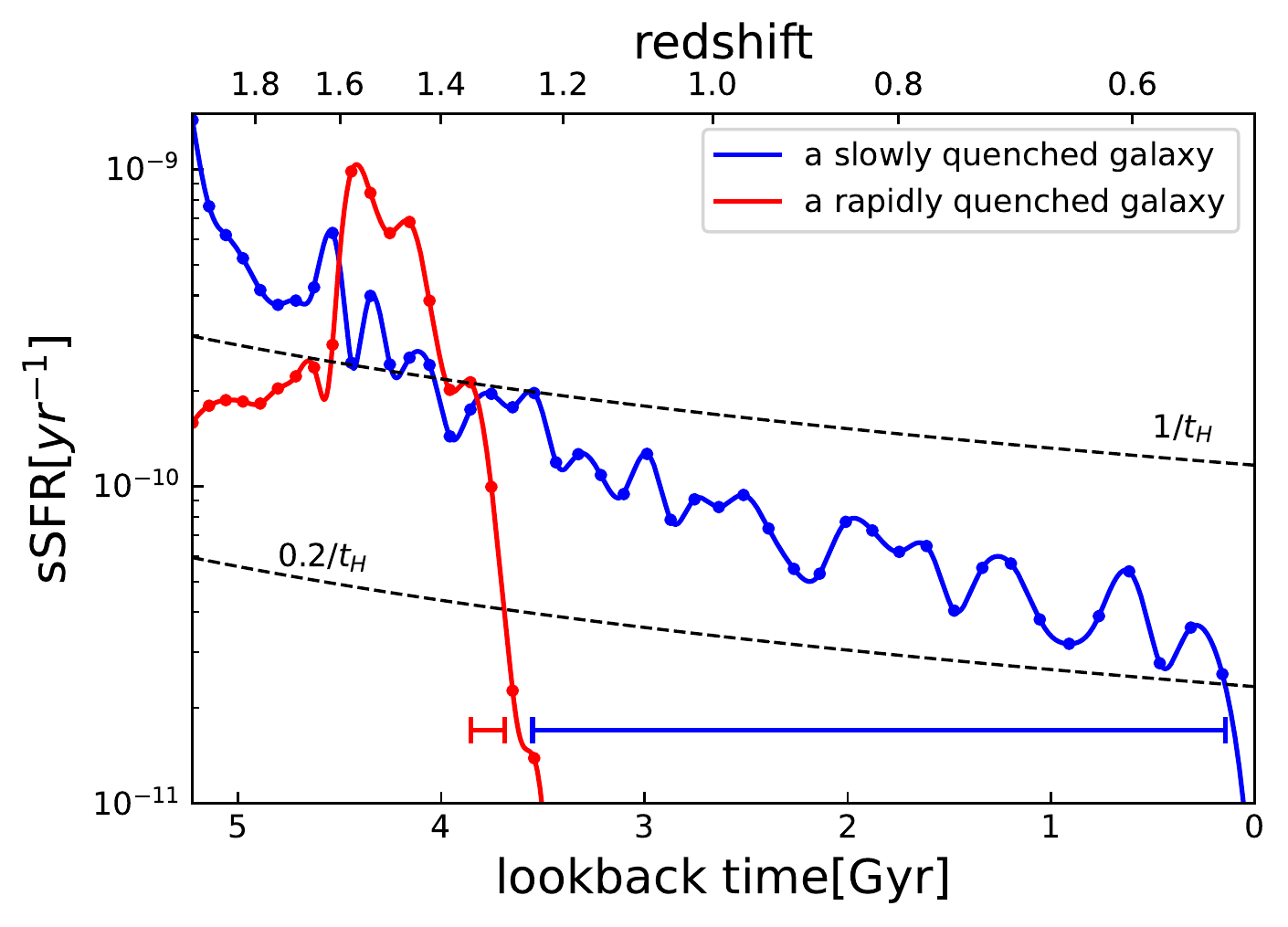}
	\caption{
	Specific star formation histories of an example rapidly quenched galaxy (red dots) and slowly quenched galaxy (blue dots) identified in the $z=0.5$ simulation snapshot, with lookback time from $z=0.5$.
	\revision{The dashed lines represent the star-forming (upper, sSFR($z$) $> 1/t_H(z)$) and quenched (lower, sSFR($z$) $< 0.2/t_H(z)$) thresholds.}
	To better estimate when the galaxies cross the thresholds,  we first fit a cubic B-spline \citep{dierckx1975algorithm} to the specific star formation history, which are shown as solid lines. We measure the time that a galaxy takes to cross from the star-forming to the quenched threshold and define this to be the quenching time $\tau_q$ of the galaxy, indicated by the horizontal bars in the lower portion of the plot.
	}
	\label{fig:SFH}
\end{figure}

Now that we have the full progenitor histories for the galaxies, the specific SFHs (i.e. sSFR vs. time) are formed from the instantaneous gas SFRs of the progenitor at each snapshot, normalised by the total stellar mass of the progenitor, against the cosmic ages of the snapshots. \autoref{fig:SFH} shows the derived SFHs for two example galaxies that are quiescent in the $z=0.5$ snapshot, with the dots showing the instantaneous sSFRs in each snapshot. \simba\ snapshots have a time spacing of $\sim100$\,Myr in the redshift range of interest in this paper, which is relatively coarse compared to the timescale over which some galaxies can be quenched. So to improve the accuracy of the estimated quenching timescales, we fit a cubic B-spline \citep{dierckx1975algorithm} to the SFH. The interpolated SFHs for the two examples are shown in \autoref{fig:SFH} by solid lines. We define the quenching time, $\tau_q$, as the time a galaxy takes to cross from a star-forming threshold to the quiescent threshold defined above, where the star-forming threshold is defined as sSFR($z$) $> 1/t_H(z)$.  Defining the thresholds with respect to the Hubble time, as suggested by \revision{\citetalias{rodriguez-montero2019} and }\cite{Pacifici2016}, accounts for the strong evolution in typical galaxy sSFR with redshift \citep[e.g.][]{speagle2014highly}. These thresholds are shown by black dashed lines in \autoref{fig:SFH}, and the \update{inferred} $\tau_q$ of the example galaxies are indicated by the horizontal bars towards the bottom of the figure. Note that the quenching time $\tau_q$ is sometimes scaled by the age of the Universe $t_H(z)$ to give a normalised quenching time of $\tau_q/t_H(z)$. As explained below we do not follow this convention in this paper.

\revision{We note that the SFHs of galaxies that are identified from the merger trees are different from those used to calculate the spectral energy distribution (SED) of the galaxies in a given snapshot. The SEDs are computed from the ages and metallicities of the star particles present in the galaxy at the time of observation, which includes stars originating from all pre-merged components. This replicates what is done observationally, and ensures the SEDs are created from SFHs with very high time resolution and thus able to accurately produce the SEDs of RQGs. }

\subsection{Sample selection}\label{subsec:sample}

There are $\sim$ 39,000 galaxies identified by the halo finder in the $z\sim1$ snapshot. We select the galaxies that have stellar masses more than $M_* \geq 5\times 10^9$ \Msun\ and a $K$ band apparent magnitude $K < 23 $ mag.
\update{These two constraints are chosen to match the observational sample that we are going to compare to in this paper.}
With an initial mass of gas particles of $1.82\times 10^7 $\Msun, this stellar mass cut guarantees that each galaxy contains more than 275 stellar particles so that the effects of numerical fluctuations are limited. The $K$ band cut also helps to reduce numerical effects, but more importantly, it ensures that the galaxies are bright enough to be observed in our comparison observational dataset.

\begin{figure}
	\includegraphics[width=\columnwidth]{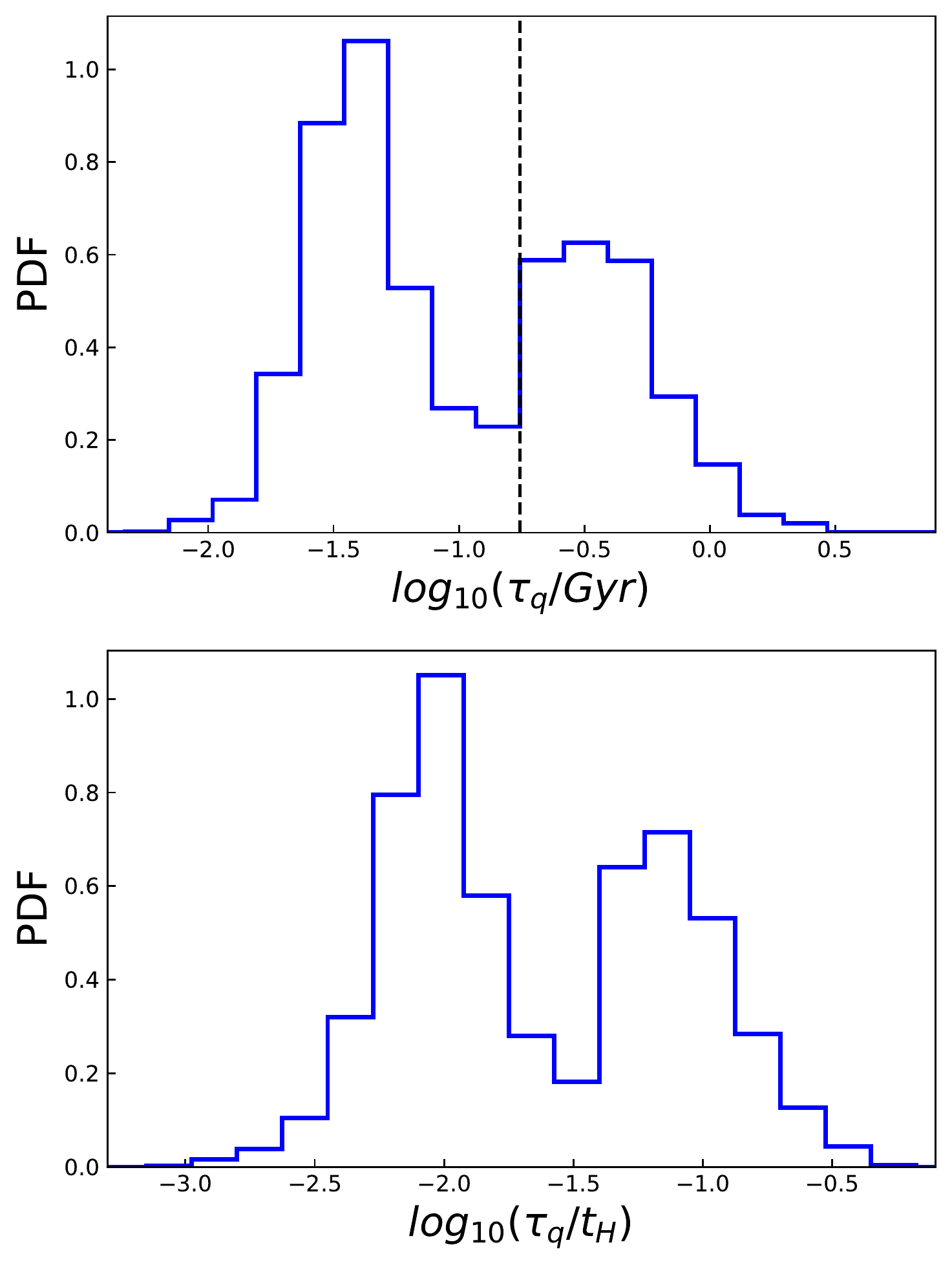}
	\caption{The probability density function (PDF) of the quenching timescale (upper panel), and normalised by the age of the Universe at the time that the galaxy is quenched (lower panel) for galaxies that are quenched by $z=1$ in \simba. In both cases the distribution is clearly bimodal, with a division at $\tau_q\sim 175$ Myr (dashed line) or $\tau_q/t_H \sim 0.03$. In this paper, we are interested in the spectral evolution of the galaxies, therefore the quenching timescale \emph{without} normalisation is the relevant property.  
	}
	\label{fig:quenching_timescale}
\end{figure}

The mass and apparent magnitude cuts select 10,530 galaxies. 
Of these, 2623 lie below the quiescent threshold (sSFR $< 0.2/t_H$) and form the red sequence at redshift $z\sim 1$. As described above, we track back the progenitors of each galaxy in this final sample, reconstruct their SFH and measure their quenching timescales. The upper panel of \autoref{fig:quenching_timescale} shows the stark bimodality in the distribution of the quenching timescales very clearly, with a division at $\tau_q\sim175$ Myr (the dashed line in the figure). For slowly quenched galaxies, their quenching timescale can be comparable to the cosmic timescale, hence it is common to normalise the times by the age of the Universe at the time that the galaxy is quenched ($\log(\tau_q/t_H)$, lower panel of \autoref{fig:quenching_timescale}). The bimodality remains with a division at $\log(\tau_q/t_H)=-1.5$ \revision{or $\tau_q/t_H\approx 0.032$ equivalently}. While this normalisation is common in the literature (see e.g. \citetalias{rodriguez-montero2019}
where the bimodality was discovered in \simba), it is unrelated to the spectral evolution of the galaxies, which evolve the same regardless of Cosmic age.  Thus in this paper, we adopt the criterion \emph{without} Universe age scaling and define ``rapidly quenched galaxies'' (RQGs) as those with $\tau_q \le 175$ Myr\footnote{\revision{We test the impact of varying this threshold to be 150 Myr or 200 Myr. The variation does not lead to any qualitative difference in our results and only changes the numbers by a few per cent.}
}.

\subsection{Mock photometric dataset}
Constraining the SFH of a galaxy with observational data is challenging, with limited information stored in the integrated spectral energy distributions (SEDs), and strong dependencies on the fitting methods and stellar synthesis models. Therefore, to compare with real data, we use the {\sc pyloser}\footnote{\url{https://pyloser.readthedocs.io/en/latest/}} package to forward-model the observed colours and create mock photometric data for the simulated galaxies in \simba. This uses the stellar population synthesis models pyFSPS, a python version of FSPS \citep{conroy2009propagation, conroy2010propagation},
which combines the MIST stellar isochrone model \citep{Dotter2016, Choi2016, Paxton2011, Paxton2013, Paxton2015} with the MILES spectral library \citep{SanchezBlazquez.etal.2006a,FalconBarroso.etal.2011a}. Throughout our analysis a Chabrier initial mass function \citep{Chabrier2003} is assumed and nebular emission is included \citep[see][for more details]{Byler2017}.
The models are interpolated to the age and metallicity of each stellar particle.

Wavelength dependent dust extinction is included by ray-tracing the line-of-sight extinction.  {\sc pyloser} calculates the dust extinction to each stellar particle within the galaxies based on the density of the line-of-sight dust column, combined with a composite dust extinction law. 
For galaxies with an average metallicity lower than one-tenth Solar metallicity, the extinction law of the Small Magellanic Cloud is applied. For galaxies with a metallicity higher than Solar metallicity, a mixed extinction law is used depending on the sSFR of the galaxies: Milky-Way extinction \citep{cardelli1989relationship} for less star-forming galaxies (sSFR $< 0.1 {\rm Gyr}^{-1}$),  \cite{calzetti2000dust} for highly star-forming galaxies (sSFR $> 1 {\rm Gyr}^{-1}$) and a linear combination of the two for galaxies with a sSFR in between. For galaxies with an intermediate metallicity that do not fall in the two groups mentioned above,  the SMC  and ``mixed'' extinction law are themselves mixed. {\sc pyloser} is much faster than full radiative transfer, however, the results are generally similar \update{(see Akins et al. in preparation)}.

The attenuated spectra of all stars in the galaxies are summed to form the integrated galaxy spectrum, which is then redshifted and dimmed according to the redshift of the galaxies (including peculiar velocities) to simulate observed spectra. Finally, filter transmission functions are applied to calculate the apparent magnitude of the galaxies in different bands. Here we compute photometry in \revision{11} filters to mimic our comparison observational dataset: the Subaru B, V, R, i', z'; UKIRT J, H, K; VISTA Y; and IRAC 3.6 and 4.5$\mu m$ bands. 

\subsection{Super-colour analysis}
\label{subsec:sca}

\begin{figure*}
	\includegraphics[width=1.6\columnwidth]{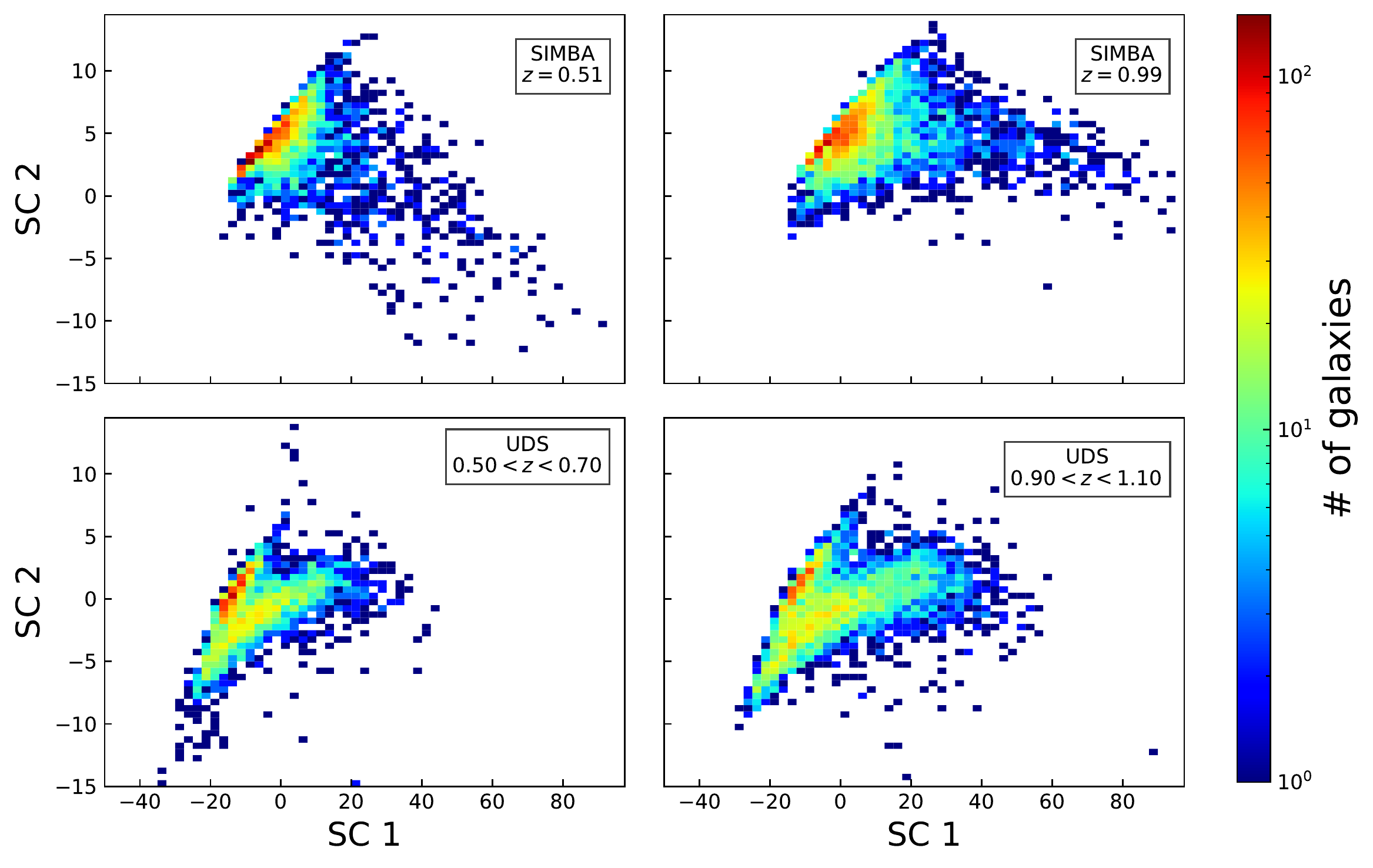}
	\includegraphics[width=1.8\columnwidth]{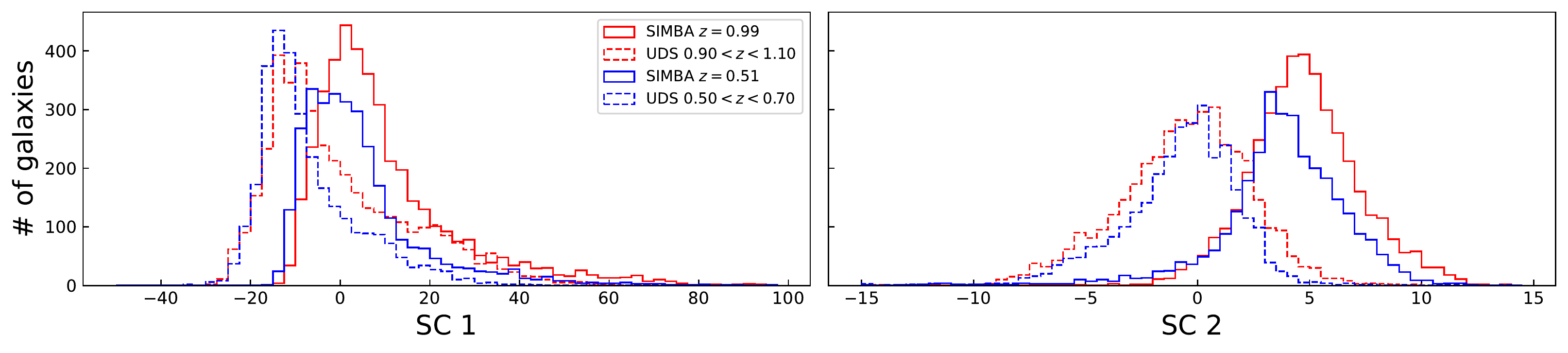}
	
	\caption{The distribution of the first two super-colours, which describe the SED shapes, of \simba\ galaxies (upper panels) and galaxies in the UKIDSS Ultra Deep Survey (lower panels). All galaxies are selected to have K-band magnitudes brighter than 23\,$mag$ and stellar masses  $M_{*} \geq 5\times10^9$\Msun. The \simba\ galaxies are selected from the $z=0.5$ (\revision{left 2D histograms}) and $z=0.99$ (\revision{right 2D histograms}) snapshots, while the UDS galaxies have a redshift range of $0.5<z<0.7$ (left) and $0.9<z<1.1$ (right). For the purposes of presentation only, in order to better compare with the observations, we plot a randomly selected 40\% and 20\% of the available \simba\ galaxies. \revision{We also show the projected 1D histograms of SC1 and SC2 in the bottom panels. }} 
	\label{fig:SCA_results}
\end{figure*}

As well as identifying rapidly quenched galaxies directly from their star formation histories, we use a second method that is more directly comparable with observations. We classify the \simba\ galaxy spectral energy distribution (SED) shapes, using a principal component analysis (PCA) of the mock observed multi-wavelength photometry.  The PCA method compresses the complicated SEDs into a few linear combinations of observed bands, called super-colours. The super-colour analysis was first introduced in \cite{Wild2014b}, and then slightly updated in \cite{wild2016} to classify galaxies in a slightly wider redshift range ($0.5<z<2$). Details of the super-colour technique can be found in these two papers, but here we provide a brief summary of the key features. 

The super-colour analysis involves projecting the multi-wavelength photometry onto pre-computed eigenvectors (or equivalently eigenspectra) that were determined from a large library of 44,000 model SEDs created by combining the \cite{bruzual2003stellar} stellar population synthesis model with `stochastic’ star formation histories \citep{kauffmann2003stellar, gallazzi2005ages} and two component dust attenuation \citep[][]{charlot2000simple}. The precise details of this library are irrelevant, as it simply defines the weighting vectors for the projection of any other dataset. This method therefore allows the direct comparison of simulated and real data, without any dependence on population synthesis models, so long as the same observed wavebands are used. The mock data is massively ``gappy'' compared to the eigenvectors which are computed using small steps in redshift in order to cover all possible combinations of observed photometry for galaxies over the full redshift range of $0.5<z<2$. This is accounted for using the ``gappy PCA'' method of \cite{connolly1999robust}, leading to formal errors on the final super-colours which depend slightly on redshift.

The principal component amplitudes indicate the contribution of each eigenspectrum in the SED and are termed ``super-colours'' (SC) as they are equivalent to a traditional colour, i.e. they are weighted linear combinations of the observed fluxes. \cite{Wild2014b} found that the first three eigenspectra are required to account for $>99.9\%$ of the variance in the model SEDs, and can thus be used to accurately, succinctly and uniquely describe and reconstruct the shape of all galaxy SEDs. The 10-band observed photometry, for galaxies over a wide redshift range, are compressed into just 3 numbers with little loss of information. Each super-colour indicates a particular property of the galaxy SEDs: SC1 describes the red-blue slope of the SED, indicating the average sSFR; SC2 changes the strength of the 4000\AA\ or  Balmer break, indicating the fraction of stellar mass formed in the last $\sim1$\,Gyr as well as the metallicity of star-forming galaxies; SC3 also controls the exact SED shape around the 4000\AA\ break and helps to break the degeneracy between metallicity and burst fraction. 

Similarly to traditional rest-frame UVJ analyses, the super-colour analysis can be used to classify galaxy SEDs into quiescent and star-forming, but without requiring model fits to determine rest-frame colours. Importantly for this project, the super-colours cleanly separate out an unusual class of rapidly quenched, recent starburst galaxies (i.e. post-starburst galaxies), which lie above the main populations due to their excess population of A- and F-type stars which have unusually strong Balmer breaks while still being quite blue \citep{wild2016, Maltby2018, Wild2020star}. For the purpose of identifying these rapidly quenching galaxies, we find that SC1 and SC2 are sufficient so we focus on these two SCs in the main body of the paper and briefly discuss SC3 in Appendix \ref{sec:discrepancies}.

The upper panels of \autoref{fig:SCA_results} show the super-colour distributions of the \simba\ galaxies at $z=0.5$ and $z=1$\footnote{For the purposes of presentation only, we select a random subset of 40\% and 20\% of the available galaxies in order to approximately match the number of galaxies in the observational comparison samples.}. In the \revision{middle} panels, we display the observed distribution of super-colours for galaxies \citep{wild2016} from Data Release 8 (DR8) of the United Kingdom Infrared Telescope (UKIRT) Ultra Deep Survey (UDS, \citealp{lawrence2007ukirt}, Almaini et al. \textit{in prep}), with the same mass and K-band magnitude cut, but with slightly looser redshift ranges\footnote{\revision{Ideally, a redshift range of $0.4<z<0.6$ would be used here, however, the UDS super-colour catalogue only includes galaxies at $z>0.5$ \citep[][]{wild2016}.}} of $0.5<z<0.7$ and $0.9<z<1.1$. In the observational data, we observe a tight sequence in the upper-left region of super-colour space, clearly separated from the lower branch, which is comprised of quiescent galaxies with low sSFR \citep{Wild2014b}. The lower branch contains star-forming galaxies spread out to cover a much larger region in SC1-SC2 space, with higher sSFR galaxies on the right and dustier or low sSFR galaxies on the left.  The post-starburst galaxies lie to the upper-right end of the red sequence, forming a tight sequence and clearly separated from the star-forming cloud, unlike in traditional UVJ colour-colour diagrams.

Comparing between the \simba\ and UDS samples in the upper and \revision{middle} panels,
\update{\simba\ broadly reproduces the important features like the red sequence and the blue cloud,  but in more detail}
we see some very clear differences, indicating that the shapes of the modelled SEDs are not entirely correct in \simba. The \simba\ galaxies are shifted to higher SC1 and SC2 values compared to the UDS, i.e. the modelled SEDs are on average too blue and have too strong Balmer and 4000\AA\ break strengths,\revision{which can be seen more clearly in the 1D distributions in the bottom panels in \autoref{fig:SCA_results}}. There is also no clean separation between the quiescent and blue sequence, and the post-starburst population is much less clearly defined. In Appendix \ref{sec:discrepancies} we show that the differences remain when a different stellar population model is used to create the mock photometric data, and also that the blue sequence is well separated from the quiescent population when dust extinction has not been applied. From these investigations we conclude that an inaccurate dust treatment may \update{in part} be causing the blue sequence to blend with the quiescent population, but the remaining differences between \simba\ and UDS hint that \simba\ is not perfect in terms of creating the right types of star formation and chemical evolution histories; we return to this below. Nevertheless, as we demonstrate conclusively below, it is still possible to use the super-colours to identify rapidly quenching galaxies, albeit with a level of contamination from star-forming galaxies that we do not expect to see in real samples.

\section{Results}
\label{sec:results}

In this section we study the effectiveness of the super-colour technique at identifying rapidly quenched galaxies, measure the contribution of the rapid quenching pathway to the build up of the quiescent population as a function of both mass and redshift, and compare the mass function with observations of galaxies selected using the same super-colour technique. Finally, we investigate the visibility timescales of the rapidly quenched galaxies and the impact of an absence of starbursts on the \simba\ galaxy colour distribution. 

\subsection{Rapidly quenched galaxies in super-colour space}
\label{subsec:box}

\begin{figure}

	\includegraphics[width=\columnwidth]{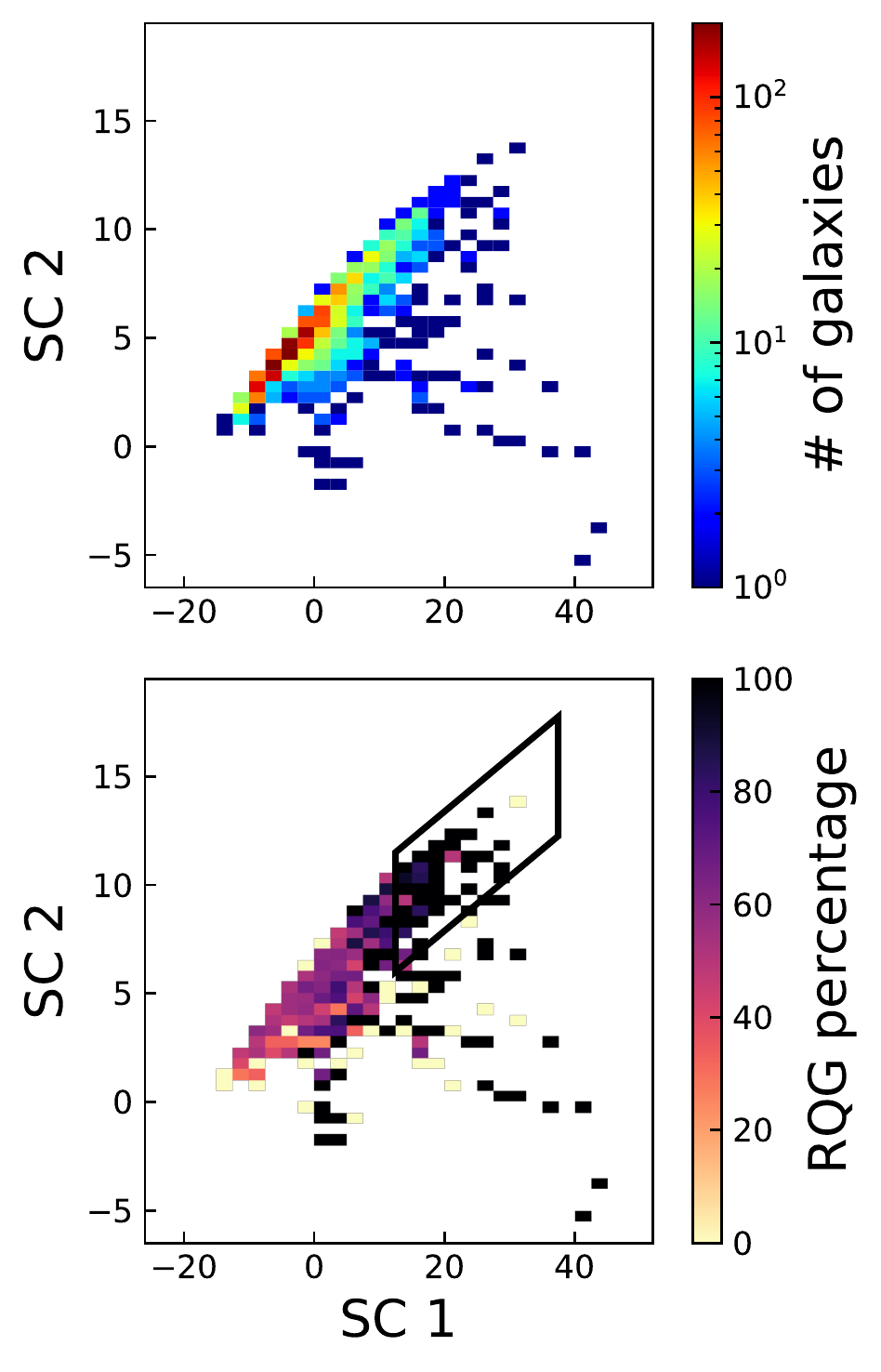}

	\caption{\textit{Top:}  The distribution in super-colour space of quiescent galaxies in the $z=1$ snapshot of the \simba\ simulation, selected to have an instantaneous sSFR$<0.2/t_H(z=1)$. As found in observed data, they form a tight sequence to the left of super-colour space. \textit{Bottom:} The ratio of the number of rapidly quenched galaxies to all quenched galaxies within each bin. Almost all quiescent galaxies found in the upper right tip of the red sequence have been rapidly quenched. We mark this region with a box to further investigate the properties of the galaxies in this region.}
	\label{fig:box}
\end{figure}

The upper panel of \autoref{fig:box} shows the distribution in SC1-SC2 super-colour space of the 2623 quiescent galaxies with instantaneous sSFR below the quiescent threshold of sSFR$< 0.2/t_H(z=1)$ in the $z=1$ snapshot (\autoref{subsec:sample}).  As expected from the observational data, the quiescent population follows a tight sequence towards the left-hand edge of the full SC distribution.  This is due to the strong 4000\AA\ break strength, with the slope in SC1-SC2 caused by the distribution of both ages and metallicities in the population. The lower panel shows the ratio of rapidly quenched galaxies ($\tau_q \le $175\,Myr) to all quiescent galaxies in super-colour space.  We  see that almost all quiescent galaxies in the upper right tip of this sequence have been rapidly quenched, matching the region identified as being dominated by post-starburst galaxies in observations \citep[][]{Wild2014b}.
We delineate the region with a box as shown to further investigate the properties of galaxies in this region of super-colour space\footnote{We note that reasonable variations in the borders of the box do not lead to a qualitative difference in our results.}. \revision{Not all RQGs fall in this region, as over time the Balmer break strength of RQGs weakens and they descend in SC1 and SC2 into the red sequence. Therefore, this box is designed only to identify RQGs that quenched recently, as we explore further below.}

\begin{figure}

	\includegraphics[width=\columnwidth]{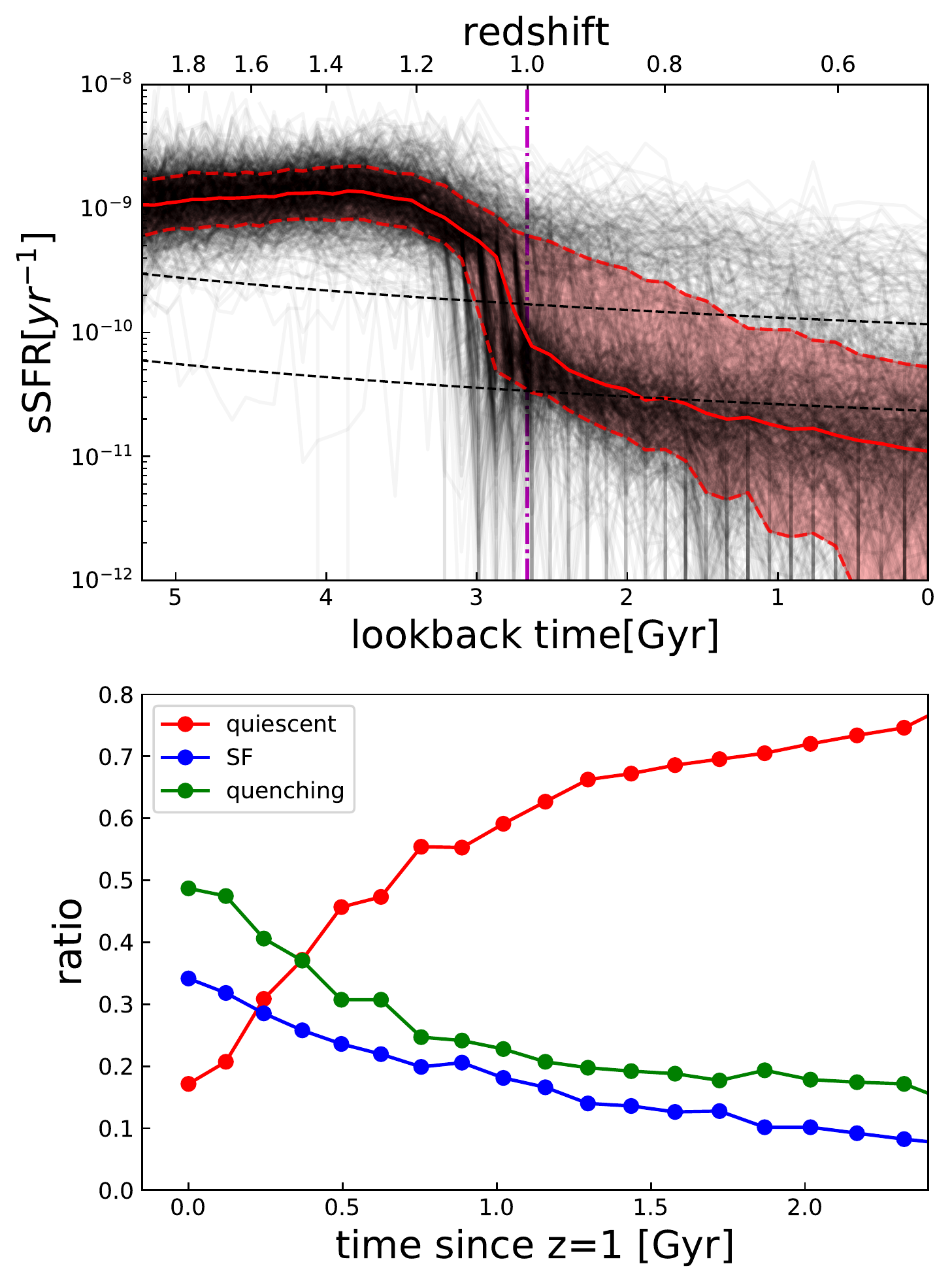}
	\caption{Investigating the past and future star-formation properties of galaxies that fall inside the box delineated in \autoref{fig:box} in the $z=1$ snapshot of the \simba\ simulation.
	\textit{Upper panel}: the sSFH of all galaxies that fall inside the super-colour defined RQG box in the $z=1$ snapshot (black lines), with the 16th, 50th and 84th percentiles of the sSFR at each lookback time since $z=0.5$ (red lines). The dashed black lines show the evolving star-forming (the upper one) and quiescent thresholds (the lower one). The magenta line marks the time at $z=1$. Although only $\sim10$\% of the galaxies in this region of SC space are currently quiescent, the majority show a recent sharp drop in their SFHs which drives their unusual SED shapes and position in super-colour space. \textit{Lower panel}: The census and future evolution of galaxies that fall in the super-colour defined RQG box at $z=1$. The majority of galaxies in the box are ``quenching'', i.e. they have a sSFR between the star-forming and quiescent thresholds. Despite their recent rapid quenching events, many remain above the quenched threshold for a significant time period, with only $\sim$60\% of the galaxies in the box at $z=1$ becoming formally quiescent in the following 1.0\,Gyr.
	}
	
	\label{fig:box_info}
\end{figure}

In the upper panel of \autoref{fig:box_info} we plot the past and future sSFH of all 729 galaxies that are found within the box in the $z=1$ snapshot (i.e. including galaxies that are not yet quiescent, unlike in \autoref{fig:box}). We find that only 10\% of them are formally quenched by our definition (i.e. have dropped below the quiescent threshold of sSFR$< 0.2/t_H$), however the galaxies have experienced a rapid drop in sSFR in the recent $<0.5$\,Gyr, as expected given their strong Balmer break strengths. The median, 16th and 84th percentiles also show a dramatic decrease of sSFR. This highlights an important point about the identification of RQGs: the spectral features are transient, and therefore they are only identifiable for a short time following quenching (see \autoref{sec:visibility}). 

In the lower panel  of \autoref{fig:box_info} we show the fraction of these galaxies with sSFR above the star-forming line \update{(blue line)}, below the quenched line \update{(red line)}, and in \update{ the process of becoming quenched (green line)}, 
as a function of time from $z=1$. Despite most having experienced a recent sharp decline in star formation, they do not immediately cross the threshold to become fully quenched, with only $\sim$60\% of the galaxies formally quiescent within 1\,Gyr.
\update{This shows that while \simba\ produces a population of galaxies that rapidly decrease in star formation rate, the mechanisms responsible are not able to entirely quench the star formation.}
We note that this is unlikely to be representative of the situation in the real Universe 
where the clear separation of RQGs from star-forming galaxies in super-colour space, as well as evidence from spectroscopy, indicates that most RQGs have both recently and rapidly decreased their star formation and entirely halted any ongoing star formation. Further exploration of any residual level of star formation in recently, rapidly quenched galaxies may provide useful constraints on feedback mechanisms. We return to this point in \ref{subsec:lack_psb} below.

\begin{figure}
	\includegraphics[width=\columnwidth]{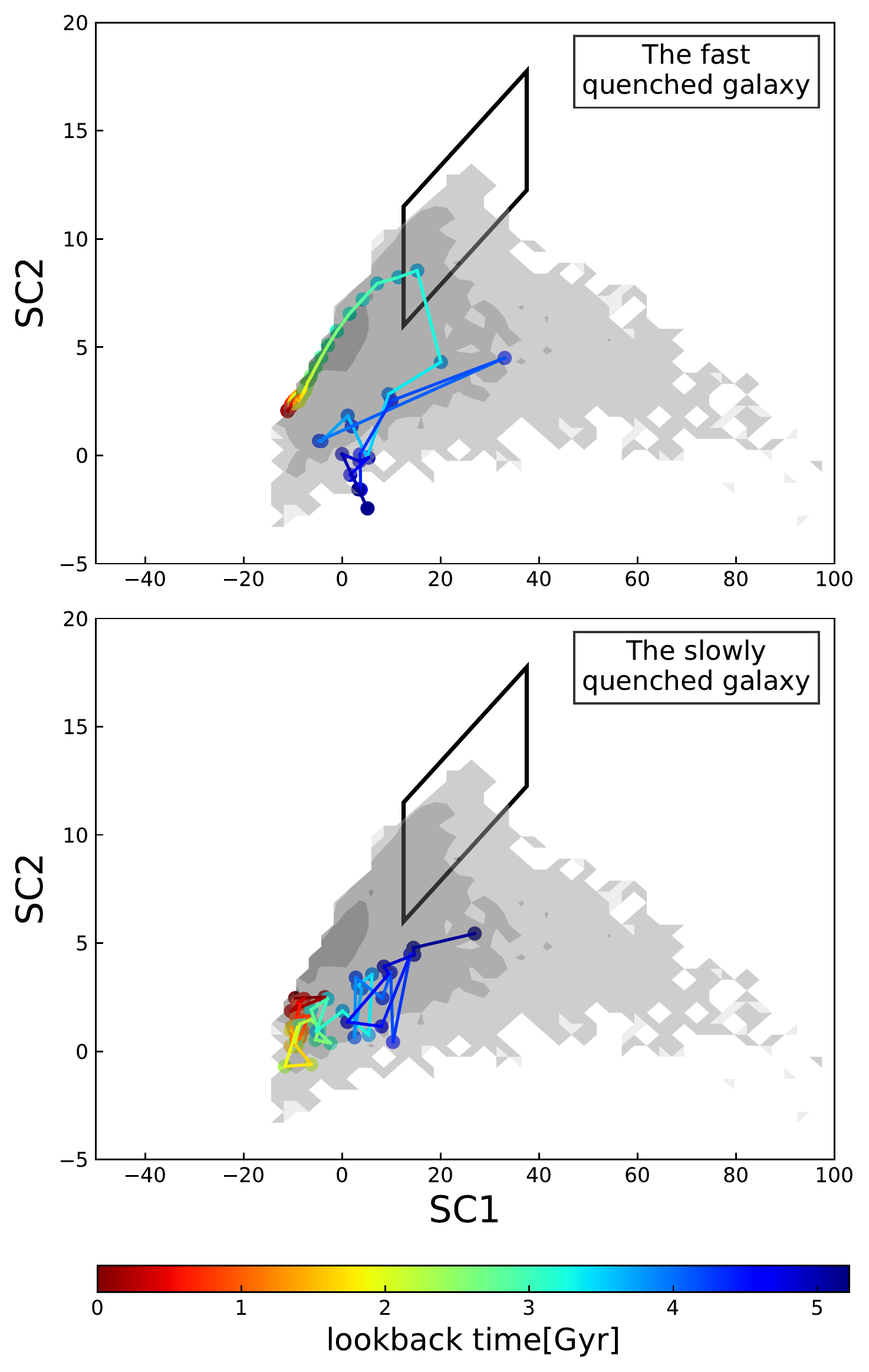}
	\caption{The evolutionary tracks in super-colour space of the rapidly quenched (upper panel) and slowly quenched (lower panel) galaxies displayed in \autoref{fig:SFH}.	The underlying grey contours show the distribution of \simba\ galaxies at redshift $z\sim1$.
	The colour bar indicates look-back time from $z=0.5$, where these two galaxies were selected to be quiescent.
	}
	\label{fig:SCA_traces}
\end{figure}

\subsection{Contribution of rapid quenching to the quiescent population}
\label{subsec:mass_func}

With our simulation dataset, there are two ways to identify quiescent galaxies that have previously rapidly quenched their star formation and thus assess the contribution of rapid quenching to the build up of the quiescent population. Firstly, we can identify them by their known SFHs (SFH-selected), and secondly by their observational signatures (SC-selected). Note that we do not use the super-colours alone to identify RQGs in \simba\ as is done observationally, but first identify quiescent galaxies and track back their evolution in super-colour space to see whether they enter the region where we find a high fraction of RQGs in \autoref{fig:box}.  In this section we compare both methods to measure the contribution of the rapid quenching route to the quiescent population in \simba, as well as how well photometric observations are able to recover the true impact of rapid quenching.

In \autoref{fig:SCA_traces}, we display the super-colour tracks of the progenitors of the two example $z=0.5$ galaxies shown in \autoref{fig:SFH}, coloured by lookback time since $z=0.5$. The upper and lower panels show the rapidly and slowly quenched galaxies respectively. While the RQG moves through the box region, the slowly quenched galaxy moves directly onto the red sequence without entering the box region. We can calculate the super-colour track of all RQGs in the $z=1$ \simba\ snapshot, identifying those that enter the box region of super-colour space, and classifying them as ``rapidly quenched'' from their observed colours\footnote{We linearly interpolate the SC1-SC2 points between individual snapshots to identify whether a galaxy enters the region, rather than using the snapshots alone.}.  
We expect this process to identify fewer RQGs than using the SFHs, which are not accessible observationally, because the super-colours are not a perfect selection method, \revision{one cannot ensure that every single RQG would enter the box region}. Because the super-colours that we are using are only defined up to a redshift of 2, it is only possible to identify ``observational'' rapid quenching events that occur after this time. In the following, we ensure that we compare like-with-like when comparing SFH-selected and colour-selected RQGs by restricting the SFH-selected RQGs to those that quench below $z_q=2$. We define $z_q$ to be the redshift at which the sSFR of the galaxy last crosses the quiescent threshold of $0.2/t_H(z)$. 

We find that 1537 out of 2623 quiescent galaxies at redshift $z\sim1$ in \simba\ have a quenching timescale of $\tau_q \le 175$ Myr, i.e. $\sim$ 59\% have been rapidly quenched. Summing their stellar masses, we find they contribute $\sim$48\% of the total stellar mass in the red sequence.
\revision{To estimate the uncertainty of the RQG fraction in \simba, we divide the \simba\ simulation box into 8 sub-volumes by the middle points of each dimension of the simulation box and analyse the RQG number ratio and mass contribution fraction to the growth of red sequence. The standard deviation of the RQG number ratio and mass contribution fraction is 3\% and 5\% respectively.}
Restricting to those galaxies with $z_q<2$ \revision{in order to better compare to the observations}, we find 1315 out of these 2314 quiescent galaxies are SFH-defined RQGs making up $\sim$57\% of the red sequence in terms of number count and $\sim$45\% in terms of stellar mass, \revision{with a standard deviation of 3\% and 6\% respectively among the 8 sub-volumes.} 
\update{Thus roughly half the red sequence at $z=1$ is comprised of rapidly quenched systems in \simba.}  We then check the tracks of the 1315 SFH-defined RQGs, and find that 1247 of them once enter the box in super-colour space, \update{showing that the SC box is an effective selector of RQGs}. Thus, 1247 out of 2314 quiescent galaxies are RQGs that can be identified by the super-colour selection, making up $\sim$54\% of the red sequence in terms of number count. However, they only account for $\sim$38\% in terms of stellar mass. \revision{The uncertainty on these two numbers is also about 3\% and 5\% respectively.}

\begin{figure}
	\includegraphics[width=\columnwidth]{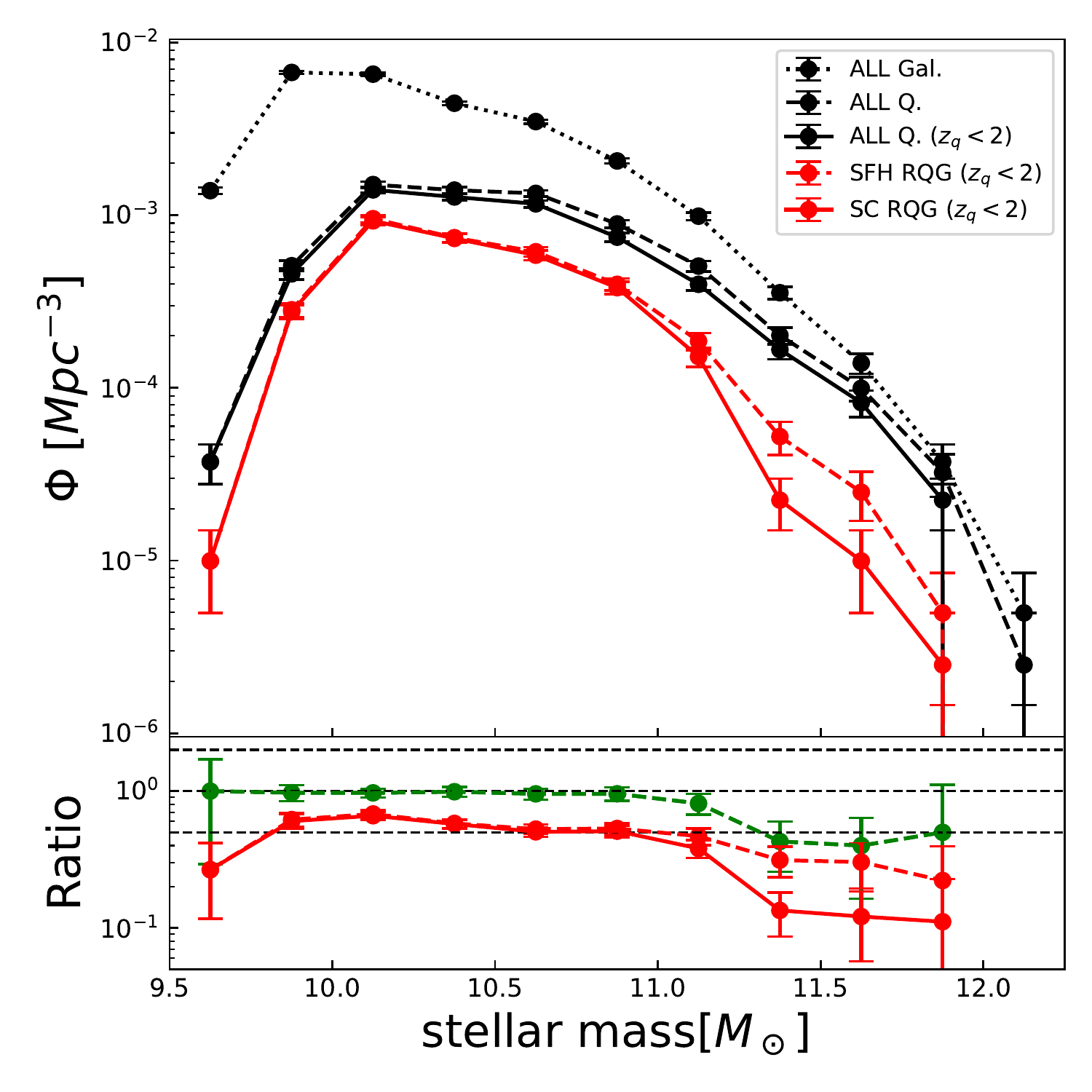}
	\caption{\textit{Upper panel}: The $z=1$ mass function of \revision{all galaxies (black dotted line, both star-forming and quiescent), }all quiescent galaxies (black dashed line, sSFR $< 0.2/t_H(z=1)$) and those quenched after redshift $z_q<2$ (black solid line); the SFH-identified rapidly quenched galaxies (red dashed line, $\tau_q\le 175$\,Myr) and the super-colour identified rapidly quenched galaxies (red solid line). The error bars here are the Poisson errors in each bin.
	\textit{Lower panel}: the ratio of the SFH-selected RQGs to all quiescent galaxies quenched after redshift $z_q=2$ (dashed red line), the super-colour selected RQGs to all quiescent galaxies quenched after redshift $z_q=2$  (red solid line) and the super-colour selected RQGs to the SFH-selected RQGs (green dashed line). The black dashed lines indicate ratios\update{, from top to bottom,} of 0.5, 1 and 2.
	}
	\label{fig:mass_function}
\end{figure}

The numbers show that the super-colour selection identifies most of the SFH-selected RQGs but preferentially misses the more massive RQGs. To investigate this further, in the upper panel of \autoref{fig:mass_function} we display the mass functions of \revision{the whole galaxy sample  at $z=1$ (both star-forming and quiescent) with a black dotted line}, the quiescent galaxy population, with a black dashed line for all quiescent galaxies, and black solid line for galaxies with $z_q < 2$, alongside those quiescent galaxies identified as rapid quenchers by their SFH (dashed red line) or from having passed through the box in super-colour space since $z<2$ (solid red line). We assume Poisson errors when we count the galaxies in each mass bin, and these are then propagated into the mass function.  The mass function peaks between $10^{10}$\Msun\ and $10^{11}$\Msun, with fewer lower and higher mass quiescent galaxies as expected. By restricting our analysis to galaxies with $z_q<2$ we lose a small fraction of quiescent galaxies, with a slight bias to preferentially losing the more massive ones. 

Clearly the shape of the mass function of the RQGs is similar to that of the total quiescent population, but to better compare the three mass functions, their ratios are plotted in the lower panel of \autoref{fig:mass_function}. The green line compares the mass function of the SFH-selected RQGs with the SC-selected RQGs, while the red lines compare the mass functions of the RQGs with the quiescent population with $z_q<2$. The dashed black lines indicate the ratios of 0.5, 1, and 2 for easy reference. We see that for the majority of the quiescent population, the rapid quenching route makes up a little over half by number count of all quiescent galaxies. The dominant difference between the shape of the mass functions is at the high mass end, with RQGs having a lower typical stellar mass than the overall quiescent population, highlighting that the rapid quenching routes are more important in the low/median mass galaxies. This coincides with the mass range where \simba's jet feedback is dominant, consistent with the idea that the star formation in these RQGs is rapidly quenched by the violent jet feedback \update{and associated X-ray feedback}. We also note an apparent drop in contribution of RQGs at the lowest masses, and it is possible that ongoing accretion prevents these galaxies from depleting their gas rapidly enough to be selected. However, the smaller number of galaxies in the lowest mass bin make this result uncertain.

From the green dashed line indicating the ratio of SC-selected RQGs to SFH-selected RQGs, we see that almost all RQGs are identified observationally at stellar masses below $10^{11}$ \Msun. At the high mass end, some SFH-selected RQGs can be missed by the super-colour identification.
We visually inspected the SFHs and super-colour tracks of the 49 RQGs with  $M_*>10^{11}$ \Msun\ that are not identified by their SC tracks. We found that they fall into two groups: 1) the galaxy does not have a high enough sSFR prior to quenching to show sufficiently strong spectral features to enter the box in super-colour space; 2) the galaxy has a fluctuating SFH after being quenched and although the sSFR never returns above the star-forming threshold, the weak rejuvenation event(s) complicate the super-colour track of the galaxy, preventing it from entering the selection box. This process appears to be particularly prevalent in the highest mass galaxies. 

\subsection{Visibility timescales}
\label{sec:visibility}
Understanding the visibility timescale of RQGs or PSBs is an active area for astronomers \citep[e.g.][]{french2021evolution}, as this allows an estimate of the mass growth of red sequence through the rapid quenching route. \revision{Measuring timescales requires high quality continuum spectroscopy to accurately reconstruct their star formation histories, and therefore their evolution through colour-selection space. Fortunately,}
for simulated RQGs in \simba, a direct measurement of the visibility timescale can be more easily obtained.

We define the visibility timescale of RQGs as the duration over which the galaxies stay in the box region of super-colour space. The time-spacing of the \simba\ snapshots is coarse relative to the visibility timescale, therefore it is not possible to do this precisely based on the snapshots only, and we instead use the linear interpolation in SC space to estimate the visibility time.  For the small fraction of tracks that start in the box, i.e. at $z=2$ the galaxy is already quenching, we simply take the time of $z=2$ as a entering time. Similarly, for tracks that end in the box, we take the time of $z=0.5$ as a leaving time. Some galaxies may enter and leave the box region several times due to fluctuations in their SFHs, and for these we accumulate all the time the galaxies super-colours are within the box. At $z=0.5$, there are 2938 quiescent galaxies that have passed through the box region with a median visibility time of \update{$\sim$ 400}\,Myr.

The visibility timescale of RQGs decreases as the redshift decreases, from \update{$\sim$ 500}\,Myr for galaxies that quenched between $1.25< z_q <2$, to \update{$\sim$ 400}\,Myr for galaxies that quenched between $0.75< z_q <1.25$ and \update{$\sim$ 300}\,Myr for galaxies that quenched at $0.5< z_q <0.75$ respectively. This difference likely arises due to the higher gas fractions and therefore higher sSFR of galaxies at higher redshifts, causing a stronger observed Balmer break feature in the SEDs which remains for longer. This is important, as it means that the observed increase in number density of PSBs at high redshift in the observations should be moderated by the increased visibility time for identifying them. 

These values are not directly comparable to observational results, due to the difference in super-colour distributions between \simba\ and the UDS meaning that different boxes were used to identify the RQGs. Therefore, the value of 390\,Myr, compared to 0.5-1\,Gyr in \citet{Wild2020star} is likely consistent within the uncertainties. This calculation should certainly be revisited when simulations are able to produce the overall colour distributions of galaxies more accurately.

\subsection{Redshift evolution}
\begin{figure*}

	\includegraphics[width=2\columnwidth]{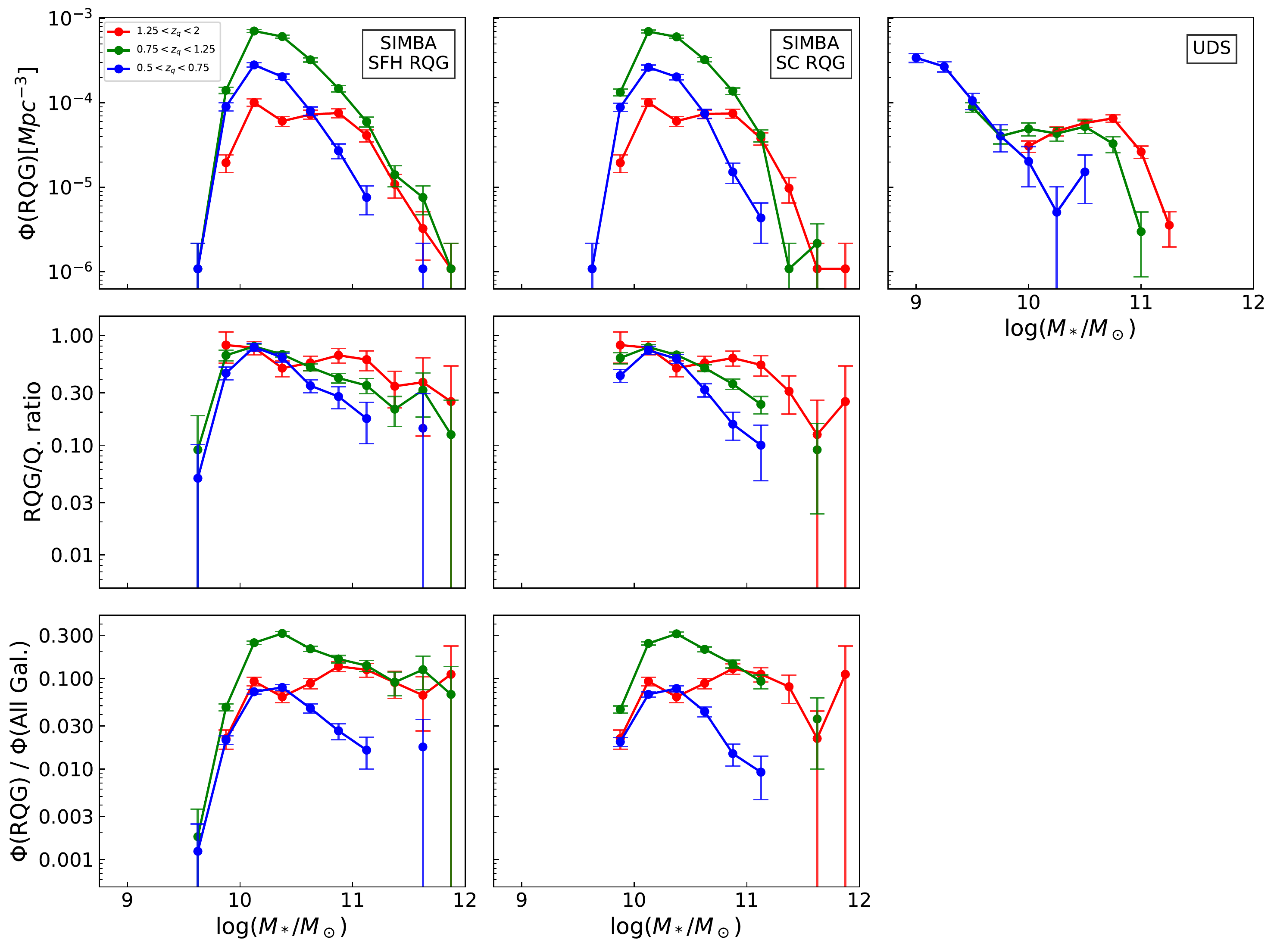}

	\caption{
	\emph{Top left:} The mass functions of RQGs identified from their SFH to have $\tau_q \le 175$\,Myr in the $z=0.5$ \simba\ snapshot, which were quenched at $1.25<z_q<2$ (red), $0.75<z_q<1.25$ (green) and $0.5<z_q<0.75$ (blue). The stellar mass is the mass at the time of quenching and the errors are propagated from Poisson errors on the number counts. \emph{Top centre:} Same as top left, but for RQGs that \emph{also} pass through the RQG region of super-colour space. \emph{Top right:} The mass functions for RQGs in the UKIDSS UDS survey, taken from the catalogue of \citep{wild2016}. 
	\textit{\revision{Middle}}: the ratio of the number of RQGs to the total number of galaxies that are quenched during each redshift interval. On the left the RQGs are identified from their SFHs alone, and in the middle they \emph{also} pass through the RQG region of super-colour space. Note that there is no easy observational comparison to these \revision{middle} panels, as it is  more challenging to identify galaxies that recently quenched slowly than to identify RQGs. 
	\revision{\textit{Bottom}: the ratio of volume density of RQGs to that of all galaxy at the middle redshift of each redshift interval. As same as the middle row, on the left the RQGs are identified from their SFHs alone, and in the middle they \emph{also} pass through the RQG region of super-colour space. }
    }
	\label{fig:MF_evo}
\end{figure*}

Observations have suggested that the importance of the rapid quenching route appears to decrease with decreasing redshift and may rapidly diminish at $z<1$ \citep{wild2016, Rowlands2018GAMA, Belli2019}. Here we investigate the redshift evolution of RQGs in the \simba\ simulated galaxies. We first select 3257 rapidly quenched \simba\ galaxies with $\tau_q \le 175$\,Myr in the $z=0.5$ snapshot, and separate them into 3 different redshift bins depending on the redshift at which the galaxy was quenched ($z_q$):  $1.25< z_q <2$ (361 galaxies), $0.75< z_q <1.25$ (1874 galaxies), and $0.5< z_q <0.75$ (650 galaxies)\footnote{The remaining 372 RQGs were quenched before $z=2$, hence are not included in this analysis.}. The mass functions of these three samples are shown in the top-left panel of \autoref{fig:MF_evo}, using the total stellar masses of the galaxies at the time they were quenched. The top-centre panel shows the same result using only those RQGs which also pass through the region of super-colour space which would allow them to be detected observationally. 

The \simba\ RQG mass functions show a strong redshift evolution, with high-mass RQGs (M$^*>10^{11}$M$_\odot$) predominantly present in the $0.75< z_q <1.25$ and $1.25< z_q <2$ samples. The number density of intermediate mass RQGs ($10^{10}<$M$^*$/M$_\odot<10^{11}$) increases rapidly between  $1.25< z_q <2$ and $0.75< z_q <1.25$. As time goes on, the mass function retains its shape but the overall number density decreases to the $0.5< z_q <0.75$ sample. The left and middle panels show very similar results, showing that the observational super-colour selection method for RQGs is able to tell a qualitatively similar story. 

The \revision{middle} panels of \autoref{fig:MF_evo} show the ratio of RQGs to the total numbers of galaxies quenched during each redshift bin in \simba, with a pure SFH selection on the left, and including the super-colour selection in the middle. Note that this only includes galaxies quenched within the redshift interval, rather than the full number of quenched galaxies at that redshift. In the high redshift bin ($1.25<z<2$) we see that the majority of galaxies that are quenched in this redshift range are quenched rapidly, at all masses. As redshift decreases, the rapid quenching pathway remains important at lower masses, but a smaller fraction of high mass galaxies pass through the rapid quenching pathway. Again, the overall qualitative trends are identified when the super-colours are used to identify the RQGs.

\revision{
The bottom panels of \autoref{fig:MF_evo} show the ratio of volume density of RQGs to that of all galaxies at a comparable redshift. Note that since $\Phi(RQG)$ is measured at the time that RQGs get quenched, each RQG is only counted once within the redshift interval. Thus for comparison, we compute $\Phi(All\ Gal.)$ at the \simba\ snapshots with the closest redshift to the middle redshift of each interval, which appropriately counts each galaxy in \simba\ once.  This assumes that the mass function does not evolve much within each redshift range, which we have verified.  The chosen snapshot redshifts are $z=1.62, 0.99, 0.63$ for the three  intervals, respectively.}

\revision{
Massive galaxies are more likely to be quiescent as seen in the upper panel of \autoref{fig:mass_function}, however the middle panels of \autoref{fig:MF_evo} it is seen that they have lower RQG fractions than the peak value at $\log\ M_*\approx 10.3$.  In the bottom panels of \autoref{fig:MF_evo} taking the ratio of RQGs to all galaxies, these two effects partially cancel each other. The former effect tends to be more pronounced, leading to a flatter trend at the high mass end compared with that seen in the middle panels.  The evolutionary trend with redshift, meanwhile, is similar to what we found in the upper panels, with RQGs peaking at $z\sim 1$.}

The top-right panel of \autoref{fig:MF_evo} shows the mass function of super-colour selected RQGs in the UKIDSS Ultra Deep Survey (UDS) for comparison, which is recreated using the data of \cite{wild2016} with the same redshift and mass bins as used in this paper. Although they are named post-starburst galaxies in the observational work, this is only a matter of differing terminology as they are selected using a similar ``box'' in super-colour space designed to catch the spur of galaxies to the upper right of the red sequence. In the $1.25< z_q <2$ redshift bin, we see a very similar shape to that seen in \simba, and there is also a similar redshift evolution at high masses. However, there are three distinct differences. Firstly, the loss of high mass RQGs clearly starts at a higher redshift in the data than in \simba. Secondly, at intermediate redshift, the number of intermediate mass RQGs ($10^{10}<$M$^*$/M$_\odot<10^{11})$) is significantly overestimated by \simba. Thirdly, in the low redshift bin the observations are dominated by low-mass RQGs (M$^*<10^{10}$M$_\odot$) which are entirely absent in \simba. The first and second discrepancy may be related, implying the jet mode feedback is operating too efficiently at $0.75<z<1.25$ causing too much rapid quenching in this redshift range. 
\revision{As for the third discrepancy, the RQGs in the UDS at low-$z$ and low mass are likely environmentally quenched \citep[][]{socolovsky2018, Wilkinson2021}, which may indicate that environmental effects are not modelled sufficiently accurately in \simba. Besides, there are also other possible reasons for the third discrepancy like 1) numerical effects caused by the limited \simba\ resolution (the galaxies at these masses have only $\sim$275 stellar particles); 2) the AGN jet feedback in \simba\ is too sub-dominant in these less massive galaxies.}
Detailed reasoning or proof for our hypotheses is difficult to obtain, hence beyond the scope of this paper.

\subsection{``Rapidly quenched'' or ``post-starburst'' galaxies?}
\label{subsec:lack_psb}

In \simba, the red sequence in super-colour space lacks the clear spur of RQGs to the upper-right, that is well separated from the star-forming cloud like in the UDS data. By fitting models to good quality spectra and multi-wavelength photometry of galaxies in the UDS field, \citet{Wild2020star} showed that these galaxies have a strong burst of star formation, hence the observationally determined name of ``post-starburst'' galaxies. \update{Given the limited resolution, \simba\ may not be able to produce the sharp bursts often invoked to represent PSBs,} and in \autoref{fig:box_info} the star formation histories do not show an obvious sharp rise prior to the quenching in most cases.

\begin{figure}

	\includegraphics[width=\columnwidth]{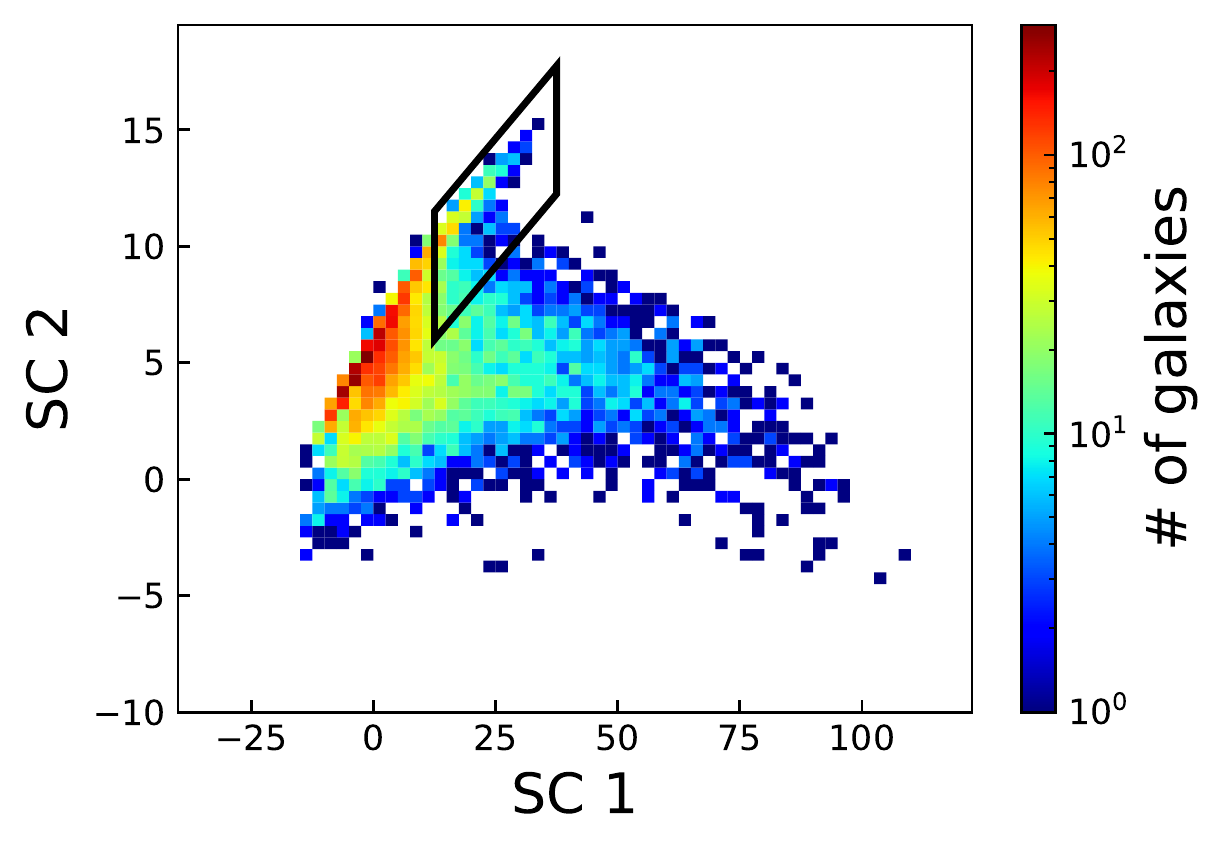}

	\caption{The super-colours of \simba\ $z=1$ galaxies with a "pseudo-burst" added (see text for details). This causes a much more prominent spur to the upper-right end of the red sequence, similar to that seen in the UDS data (\autoref{fig:SCA_results}).
    }
	\label{fig:PSB}
\end{figure}

Here we investigate the impact on the colour distribution of adding a pseudo-burst to the star formation histories of the quenching galaxies. We first select the galaxies that lie between the star-forming and quiescent thresholds, i.e. $0.2/t_H < sSFR <1/t_H$ in the $z=1$ snapshot. We then identify all newly formed stellar particles with an age $<300$ Myr and the manually change their age to 300\,Myr so that these galaxies then have a pseudo starburst 300\,Myr ago followed by complete quenching of the star formation. The stellar population synthesis process, the line-of-sight dust extinction, photometry and super-colour computation are then carried out as for the original analysis of the \simba\ galaxies, and the resulting super-colour distribution is shown in \autoref{fig:PSB}. With this pseudo-burst the thin spur to the upper-right of the red sequence is now as obvious as in UDS observations. 

Hence we conclude that the lack of this feature in \simba\ is likely a direct result of galaxies not undergoing strong enough rapid starbursts \update{prior to complete quenching}. For this reason, the term ``rapidly quenched galaxy'' is more appropriate for the \simba\ galaxies, while in the real Universe it seems that most galaxies in this region of super-colour space are in fact ``post-starburst galaxies.''  \update{The lack of direct connection between starbursts and quenching noted in \citetalias{rodriguez-montero2019} thus may indicate a failing of the model, at least for a subset of quenched galaxies.}

\section{Discussion}
\label{sec:discussion}
Unlike in \citetalias{rodriguez-montero2019}, we adopt a slightly different definition of rapid quenching: the quenching timescale is \emph{not} scaled by the Hubble time in this paper. The bimodality in quenching timescale still exists and we argue that our definition has two advantages: 1) the quenching timescale is comparable to the stellar evolution timescale of the massive stars that dominate the spectral properties, thus it is easier to link the RQGs in \simba\ with those identified by their observed spectral energy distributions; 2) this definition is more likely to relate to any physical timescale of catastrophic events that could rapidly halt star formation in galaxies, rather than slow gas-consumption type quenching mechanisms that might be expected to scale with the Hubble time. It is in part this different definition of RQGs that allows us to identify that quenching events at $z=1$ and above is preferentially rapid, a trend that is missed in \citetalias{rodriguez-montero2019}. However, both definitions guide us to the conclusion that a significant fraction of quiescent galaxies are quenched via rapid processes. 
Though observations suggest that the importance of the fast-quenching route may rapidly diminish at $z < 1$ \citep[][]{Rowlands2018GAMA}, the contribution from RQGs to the growth of the red sequence should not be ignored, at least over the first half of cosmic history. \revision{This has important implications for the interpretation of results within the current paradigm of mass vs. environment quenching \citep[e.g.][]{peng2010mass}. }

We still do not have a clear idea about what mechanisms are responsible for the rapid quenching. Both \citetalias{rodriguez-montero2019} and our results find that the rapid quenching route is important for galaxies with stellar masses around $10^{10}$--$10^{11}$\Msun. This mass range coincides with where supermassive black hole driven jet feedback is effective in \simba, suggesting that the fast quenching is likely associated with the violent expulsion of gas by the jets. \update{In \simba, however, the jets explicitly do not carry material out of the ISM as they are briefly hydrodynamically decoupled, so it is interesting that it nonetheless rapidly quenches galaxies; the X-ray feedback which pushes gas outwards and provides inside-out quenching may play a role~\citep{Appleby2020}.}

In \citeauthor{zheng2020comparison}'s
binary merger simulations, it is confirmed that AGN feedback is necessary to fully shut down the star formation in post-merger, post-starburst galaxies. However, they also find that the star formation in starburst galaxies can be rapidly suppressed to a ``main-sequence'' level without any contribution from black hole feedback, which supports the picture that stellar feedback can also play an important role in the rapid quenching route, at least in the early post-starburst phase. The stellar feedback helps driving gas outflows and disrupting the giant molecular clouds, which contributes to the end of the starburst. The exact role of stellar feedback in quenching remains unclear, and more discussion can be found in \cite{french2021evolution}.
As \simba\ does not reproduce enough starburst galaxies (see \autoref{subsec:lack_psb}), the contribution from stellar feedback to the rapid quenching will not be fully included, leading to a potential underestimation in the importance of fast quenching route, or need for enhanced AGN feedback to be included. It would certainly be valuable to further investigate the mechanisms that are responsible for the fast quenching route in \simba. Investigating other cosmological simulations such as Illustris-TNG \citep{springel2018first} and EAGLE \citep[][]{schaye2015eagle} could also help answer the question.

Mergers are believed to be an important route for rapid evolution to early-type galaxies through a starburst phase. Observationally, in the local Universe at least, many PSBs/RQGs are found to have tidal features. The consistency between the observed rates of PSBs and the merger rates at $z\lesssim1$ also suggests a predominantly merger origin of rapid quenching \citep[][]{snyder2011k+}. However, \citetalias{rodriguez-montero2019} find a different picture in \simba: not enough mergers are spotted to explain the number of quenched massive galaxies at $z\lesssim1.5$. Furthermore, they find that galaxy quenching has no obvious correlation with major mergers in \simba, disconnecting the rapid quenching route from merger induced starbursts \citep[see also][for a similar result in Illustris-TNG]{Quai2021}. Clearly further work is required to understand these apparently disparate results, both observationally and theoretically. 

Observationally the rapidly quenched satellites in clusters are found near to cluster centres \citep[][]{owers2019sami}, and it is now well established that the fraction of PSBs steadily increases from 1\% in field to $\sim$ 15\% in dense clusters \citep[][]{socolovsky2018,paccagnella2019strong}.
Unlike galaxies in the field, galaxies in a dense environment are expected to experience a much ``harder life": ram pressure stripping, galaxy harassment, thermal evaporation, and starvation. These additional processes are expected to help to rapidly stop the star formation and result RQGs. In \simba\ it is found that central galaxies are more likely to be rapidly quenched compared with satellites galaxies at the same stellar mass \citep{rodriguez-montero2019}. Thus, we can expect that the RQGs identified in this paper are more likely to be central galaxies, rather than galaxies that have recently entered a large overdensity. Further comparison of the environments of RQGs in observations and simulations may help identify further similarities and differences between the real and simulated universes.

\section{Summary}
\label{sec:summary}
We have investigated rapidly quenched galaxies (RQGs) in the \simba\ cosmological simulation at $z=0.5-2$ and compared them to
\update{observations in the UKIDSS ultra-deep survey}
using super-colours from \citet{wild2016}. We track the progenitors of the quiescent galaxies to obtain their star formation histories (SFHs), from which we measure the quenching timescale $\tau_q$, the time that the galaxies take to cross from a star-forming threshold to the quiescent threshold. We confirm a bimodal distribution of quenching timescales as in \citetalias{rodriguez-montero2019}, and divide the quiescent galaxies into rapidly quenched (RQGs) and slowly quenched. Mock photometric data are generated and a super-colour analysis is carried out in order to compare the simulation to observations. 

The findings of this paper are summarised below:
\begin{enumerate}
    \item The quenching timescales of quiescent galaxies in \simba\ show a stark bimodality with the division at $\tau_q=175$ Myr, dividing the quiescent galaxies into rapidly quenched galaxies and slowly quenched ones; this echoes the results of \citetalias{rodriguez-montero2019} who found a bimodality at $\tau_q\approx 0.03t_H$.
    \item The quiescent galaxies in \simba\ form a tight sequence in  super-colour space, similar to that observed observationally, although  offset in exact positioning. 
    \item In \simba, the blue cloud blends with the quiescent population in  super-colour space unlike the observed blue sequence, which is likely due to an inaccurate dust treatment. Additionally, the discrepancies between the super-colours of galaxies in \simba\ and the observational UDS dataset hint that \simba\ is not perfect in terms of creating the right types of star formation and metallicity histories.
    \item The super-colour distribution additionally shows that \simba\ lacks galaxies that have undergone strong  rapid starbursts, thus the red sequence in super-colour space lacks the clear spur of RQGs to the upper-right that is seen in observations. 
    \item 95\% of \simba\ RQGs pass through the upper-right region of super-colour space observationally identified as containing post-starburst galaxies, proving that the super-colour method is an effective selection technique to identify RQGs. However, there is some mass dependence, with a smaller fraction identified via their super-colours at stellar masses above $10^{11}$ \Msun.
    \item We find that 59\% of the quiescent galaxies at $z=1$ in \simba\ have been rapidly quenched with $\tau_q<175$ Myr, contributing $\sim$48\% of the stellar mass growth of the red sequence. This is at the upper end of the $\sim$25-50\% derived observationally by \cite{Wild2020star} at $0.5<z_q<2$, and higher than that estimated by \citet{Belli2019} at $z\sim1.4$.
    \item RQGs have a lower typical stellar mass than the overall quiescent population, with a distinct reduction in the importance of the rapid quenching route below $10^{11}$\Msun. This coincides with the mass range where \simba's jet feedback is dominant, consistent with the idea that the star formation in these RQGs is rapidly quenched by the violent jet feedback.
    \item At $z= 0.5$, the RQGs that have passed through the box region of super-colour space have a median visibility time of \update{$\sim$ 400}\,Myr, which is a little short compared to the 0.5-1\,Gyr estimated observationally by \cite{Wild2020star}. However, the difference may be caused by the differing boundaries necessitated by the shift in super-colours between observations and simulation. Interestingly, the visibility timescale of RQGs decreases as redshift decreases, a result which requires observational verification with higher redshift spectroscopic samples. 
    \item In \simba\ the importance of the rapid quenching route decreases with decreasing redshift for galaxies with stellar masses $>10^{10}$\Msun, but remains important for galaxies with lower stellar mass even at $z=0.5$. However, the mass functions of RQGs do not compare perfectly with observations, with a significant excess of intermediate and high mass RQGs at $1<z_q<1.5$ and lack of low mass RQGs at $0.5<z_q<1$ in \simba\ compared to observations. This implies that the jet mode feedback is operating too efficiently at $0.75<z<1.25$ causing too much rapid quenching in this redshift range.
\end{enumerate}

Clearly much more remains to be understood about the rapidly quenched galaxies. The \simba\ simulation helps us to investigate the importance of the rapid quenching pathway and its evolution at different stellar mass and different redshift. Exactly which mechanism(s) is/are responsible for the rapid quenching remains as intriguing problems to be solved in the future. The discrepancies between the properties of simulated RQGs in \simba\ and real RQGs in observations suggest some flaws in \simba, which will help guide future improvements.  \update{These sorts of detailed comparisons in the observational plane can thus shed new light on the physics of galaxy formation and particularly the pathways to galaxy quenching.}


\section*{Acknowledgements}
\revision{The authors would like to thank Omar Almaini, Sandro Tacchella and the anonymous reviewer for helpful discussions and suggestions.}
YZ acknowledges support of a China Scholarship Council - University of St Andrews Scholarship.
RD acknowledges support from the Wolfson Research Merit Award program of the UK Royal Society.
VW acknowledges support from the Science and Technology Facilities Council. 
FRM is supported by the Wolfson Harrison UK Research Council Physics Scholarship.
\simba\ was run on the DiRAC@Durham facility managed by the Institute for Computational Cosmology on behalf of the STFC DiRAC HPC Facility. The equipment was funded by BEIS (Department for Business, Energy \& Industrial Strategy) capital funding via STFC capital grants ST/P002293/1, ST/R002371/1 and ST/S002502/1, Durham University and STFC operations grant ST/R000832/1. DiRAC is part of the National e-Infrastructure. 

\section*{Data Availability}
The related data of the sample galaxies are available at \url{https://doi.org/10.17630/3d5bbf39-fe0a-4997-987c-d2407f791118}, including the SFHs, the super-colours, and the alternative super-colour data computed with pseudo-bursts or with BC03 SPS model. 
The simulation data underlying this work are publicly available at \url{https://simba.roe.ac.uk}, including both raw particle information and extracted galaxy catalogues with photometry.
Other data underlying this article are publicly available from the web as listed in the footnote.



\bibliographystyle{mnras}
\bibliography{ref} 




\appendix

\section{Discrepancies between the distribution of \simba\ and UDS galaxies in super-colour space}
\label{sec:discrepancies}
\begin{figure*}
	
	\includegraphics[width=2\columnwidth]{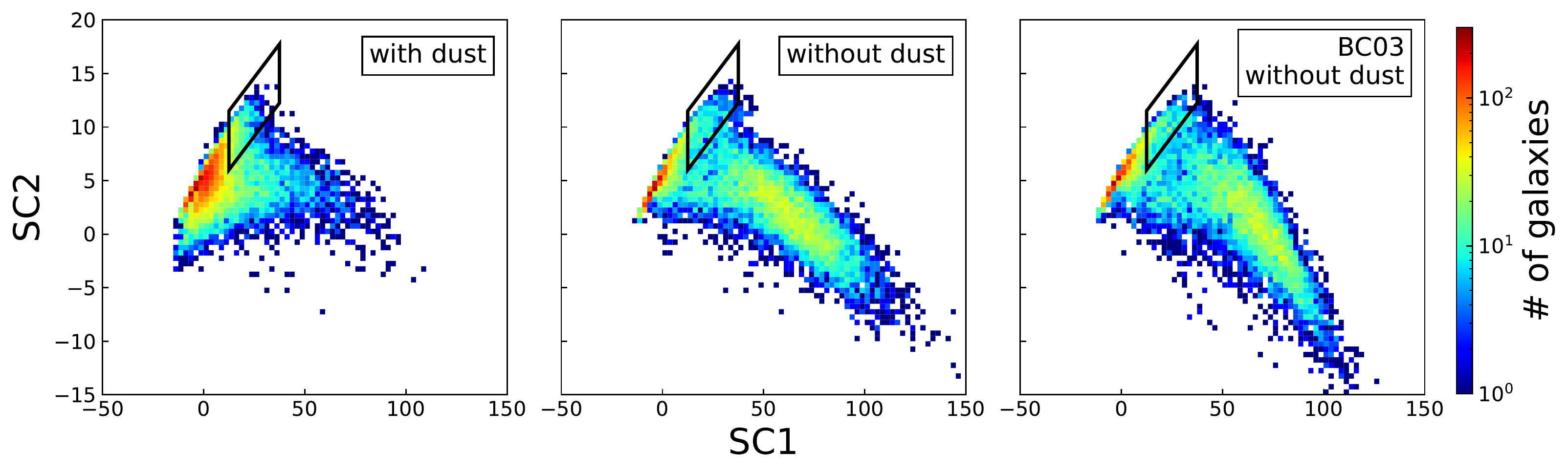}
	
	\caption{The distribution of \simba\ galaxies at redshift $z=1$ in SC1-SC2 space, the RQG selection box is plotted for reference. \textit{Left panel}: Super-colours computed \emph{including} the line-of-sight dust extinction; \textit{Middle panel}: Super-colours computed \textbf{without} dust extinction; \textit{Right panel}: Super-colours computed with the BC03 model \citep[][]{bruzual2003stellar} and \textbf{without} dust extinction.
    }

	\label{fig:diff_sca}
\end{figure*}
As noted in the main text, there are significant discrepancies between the distribution of \simba\ and UDS galaxies in super-colour space (see \autoref{fig:SCA_results})\footnote{This is \emph{not} due to the spacing of the snapshots in \simba, as the photometry is computed from the full stellar age-metallicity distribution in each galaxy, combined with line-of-sight dust extinction}. Here we try to interpret the origin of these discrepancies.

The primary discrepancy between \simba\ and UDS galaxies is the relative position of the star-forming blue-sequence to the quiescent galaxies. Unlike the galaxies in the UDS, these two galaxy populations in \simba\ do not clearly separate from each other, hence the overall galaxy distribution of \simba\ does not have a well defined ``green-valley'' region.  Star-forming galaxies have more dust than quiescent galaxies, which leads to stronger flux attenuation particularly in the blue, which can impact the measured super-colours. In \autoref{fig:diff_sca} we show the distribution of the first \textbf{two} super-colours with and without dust attenuation in the left and central columns. The blue sequence is well separated from the quiescent population when dust attenuation is excluded. It seems plausible that an inaccurate dust treatment could be the cause for moving the blue sequence too far to the red to blend with the quiescent population. 

Another suspected cause of the discrepancies is the stellar population model used when creating the mock photometric data for \simba\ galaxies. Thus, we recreate the mock photometric data for \simba\ galaxies with another stellar population model, the BC03 model \citep[][]{bruzual2003stellar}. The differences between \simba\ and UDS results remain as shown in \autoref{fig:diff_sca}

\section{Brief discussion about SC3}
\label{sec:sc3}
\begin{figure*}
	\includegraphics[width=1.6\columnwidth]{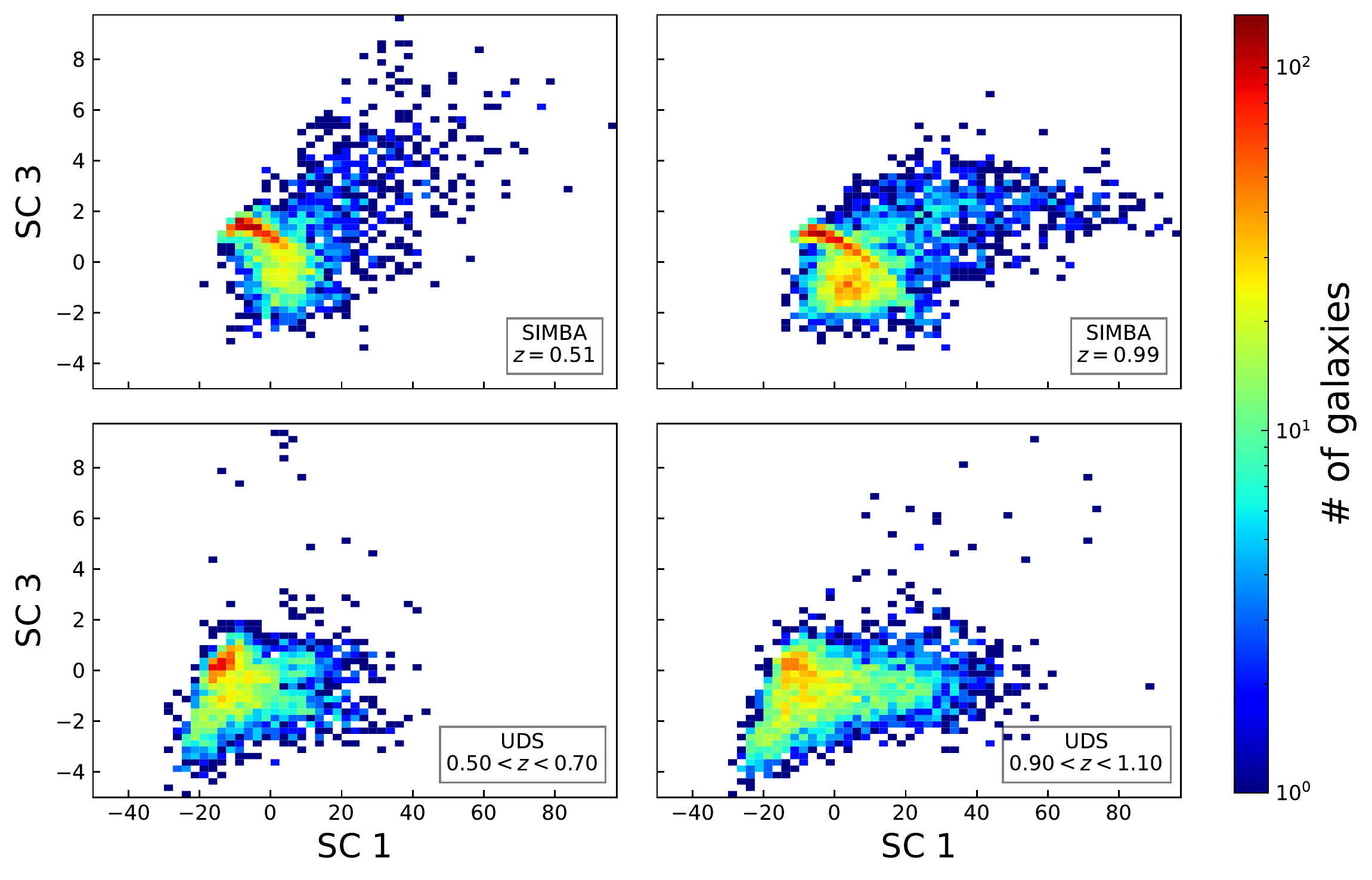}
	\includegraphics[width=0.8\columnwidth]{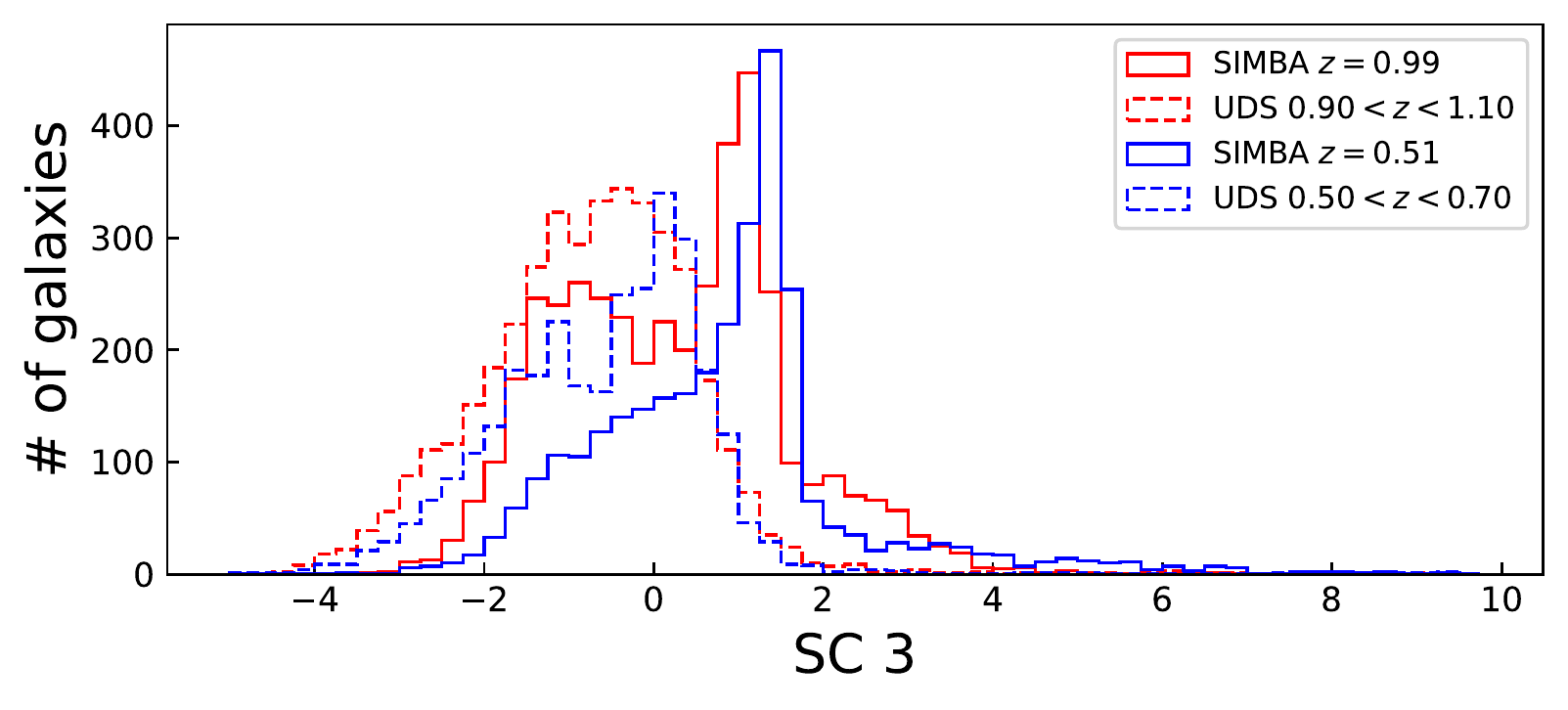}
	\caption{Same as \autoref{fig:SCA_results} but for the distribution in SC1-SC3 space.}
	\label{fig:SC13_results}
\end{figure*}

Both the second and third super-colours (SC2 and SC3) control the exact SED shape around the 4000\AA\ break, they correlate with metallicity and give additional information on stellar ages and dust.
Thus, SC3 can help break the degeneracy between metallicity, age and dust in some cases. In \autoref{fig:SC13_results}, we display the distribution of the SC1-SC3 of \simba\ galaxies (upper panels) and galaxies in the UDS (middle panels) \revision{as well as the 1D distribution of SC3 in the bottom panel}. The distribution of \simba\ galaxies show a bar structure in the SC1-SC3 space, which is clearer at $z=0.99$ (the upper right panel). We confirmed that almost all quiescent galaxies fall on this bar structure while the star-forming galaxies occupy a larger region and form the cloud.
In contrast, the UDS galaxies do not show a clear bar. However, this may be due to the measurement error on SC3 in the observations. Further investigation with only those observations where spectroscopic redshifts are available might be worthwhile to find out whether SC3 could be useful in the future.


\bsp	
\label{lastpage}
\end{document}